\documentclass{article}

\usepackage{microtype}
\usepackage{graphicx}
\usepackage{booktabs} 

\usepackage{hyperref}

\usepackage[arxiv]{icml2024}

\usepackage{amsmath}
\usepackage{amssymb}
\usepackage{bm}
\usepackage{mathtools}
\usepackage{amsthm}

\usepackage[capitalize,noabbrev]{cleveref}

\theoremstyle{plain}

\theoremstyle{definition}

\theoremstyle{remark}

\usepackage{amsmath,amsfonts,bm}

\def\eqref#1{equation~\ref{#1}}

\def\1{\bm{1}}

\DeclareMathAlphabet{\mathsfit}{\encodingdefault}{\sfdefault}{m}{sl}
\SetMathAlphabet{\mathsfit}{bold}{\encodingdefault}{\sfdefault}{bx}{n}

\usepackage[colorinlistoftodos]{todonotes}
\usepackage{ifthen}

\usepackage{titletoc}

\usepackage{listings}
\lstnewenvironment{rsverbatim}[1][]{%
  \lstset{
    basicstyle=\small\ttfamily,
    columns=flexible,
    breaklines=true,
    frame=tb,
    #1
  }%
}{}

\usepackage{soul}

\definecolor{MidnightBlue}{RGB}{25,25,112}
\definecolor{MidnightBlueComplementingGreen}{RGB}{25,112,25}
\definecolor{MidnightBlueComplementingPurple}{RGB}{112,25,112}
\definecolor{amaranth}{rgb}{0.9, 0.17, 0.31}
\definecolor{MidnightBlueComplementingRed}{RGB}{112,25,69}
\definecolor{coolblack}{rgb}{0.0, 0.18, 0.39}

\definecolor{deepjunglegreen}{rgb}{0.0, 0.29, 0.29}
\definecolor{applegreen}{rgb}{0.55, 0.71, 0.0}
\definecolor{WowColor}{rgb}{.75,0,.75}
\definecolor{MildlyAlarming}{rgb}{0.85,0.25,0.1}
\definecolor{SubtleColor}{rgb}{0,0,.50}
\definecolor{SubtleColor2}{rgb}{0.6,0.21,.50}

\definecolor{lasallegreen}{rgb}{0.03, 0.47, 0.19}

\newcounter{margincounter}

\NewDocumentCommand{\AK}{mo}{
    \IfValueF{#2}{
        {{\scriptsize
            \textcolor{violet}{ 
            \textbf{A:}
            \textit{{#1}}
            }
        }}
    }
    \IfValueT{#2}{
        \marginnote{{\scriptsize
            \textcolor{violet}{ 
            \textbf{A:}
            \textit{{#1}}
            }
        }}
    }
}

\definecolor{darkgreen}{rgb}{0.0, 0.2, 0.13}

\usepackage{tikz}
\usetikzlibrary{matrix,chains,positioning,arrows, calc}
\usepackage{float}
\usepackage{amsthm}
\usepackage{multirow}
\usepackage{multicol}

\usepackage[framemethod=TikZ]{mdframed}
\newcounter{defn}[section] 
\renewcommand{\thedefn}{\arabic{section}.\arabic{defn}}

\newcounter{theo}[section] 
\renewcommand{\thetheo}{\arabic{section}.\arabic{theo}}

\newcounter{lem}[section] 
\renewcommand{\thelem}{\arabic{lem}}

\newcounter{prf}[section]
\renewcommand{\theprf}{\arabic{section}.\arabic{prf}}

\theoremstyle{remark}

\theoremstyle{definition}

\usepackage{hyperref}       
\usepackage{url}            
\usepackage{booktabs}       
\usepackage{amsmath, amsfonts, bm}       
\usepackage{mathtools}
\usepackage{nicefrac}       
\usepackage{microtype}      
\usepackage{epigraph}
\usepackage{array}
\usepackage[normalem]{ulem}	

\usepackage{caption}        
\usepackage{sidecap}        
\usepackage{subcaption}
\usepackage{graphicx}
\graphicspath{{figs/}}
\usepackage{blindtext}
\usepackage{setspace}
\usepackage{lipsum}
\usepackage{multirow}
\usepackage[toc,page]{appendix}	
\usepackage{wrapfig}
\usepackage[bottom]{footmisc}
\usepackage{color,soul}
\usepackage{enumitem}

\definecolor{Gray}{RGB}{217,234,211}

\icmltitlerunning{Regime Adaptive Execution with Informed Data and LLMs}

\begin{document}

\twocolumn[

\icmltitle{
What Teaches Robots to Walk, Teaches Them to Trade too -- 
 \\
    Regime Adaptive Execution using Informed Data and LLMs 
}



\begin{icmlauthorlist}
\icmlauthor{Raeid Saqur}{uoft,sch,sch2}
\end{icmlauthorlist}

\icmlaffiliation{uoft}{Department of Computer Science, University of Toronto}
\icmlaffiliation{sch}{Princeton University}
\icmlaffiliation{sch2}{Vector Institute for AI}

\icmlcorrespondingauthor{Raeid Saqur}{raeidsaqur@cs.toronto.edu}

\icmlkeywords{Machine Learning, Reinforcement Learning, Quantitative Finance}

\vskip 0.3in
] 



\printAffiliationsAndNotice{}  

\begin{abstract}
Machine learning techniques applied to the problem of financial market forecasting struggle with dynamic regime switching, or underlying correlation and covariance shifts in true (hidden) market variables. Drawing inspiration from the success of reinforcement learning in robotics, particularly in agile locomotion adaptation of quadruped robots to unseen terrains, we introduce an innovative approach that leverages world knowledge of pretrained LLMs (aka. `privileged information' in robotics) and dynamically adapts them using intrinsic, natural market rewards using LLM alignment technique we dub as \textit{Reinforcement Learning from Market Feedback} (\textbf{RLMF}). Strong empirical results demonstrate the efficacy of our method in adapting to regime shifts in financial markets, a challenge that has long plagued predictive models in this domain. 
The proposed algorithmic framework outperforms best-performing SOTA LLM models on the existing (FLARE) benchmark stock-movement (SM) tasks by more than 15\% improved accuracy. On the recently proposed NIFTY SM task, our adaptive policy outperforms the SOTA best performing trillion parameter models like GPT-4. The paper details the dual-phase, teacher-student architecture and implementation of our model, the empirical results obtained, and an analysis of the role of language embeddings in terms of \textit{Information Gain}.
\end{abstract}



\section{Introduction}\label{sec:intro}
Recent advances in deep learning research have significantly changed our approach to solving complex problems in many fields.
GraphCast's~\cite{graphcast_lam2022graphcast} success in weather forecasting, 
and AlphaFold's~\cite{alphafold} breakthrough success in 3D protein structure prediction are two such emblematic examples of modern machine learning (ML) successes that have supplanted or radically shifted decades old (complex, heuristic) approaches. Legged robot locomotion is another domain that has seen breakthrough improvements in zero-shot generalization of complex terrain locomotion by successful application of (model-free) reinforcement learning techniques (along with other ingredients).   

\begin{figure}[!tb]
\begin{center}
\centerline{\includegraphics[width=\columnwidth]{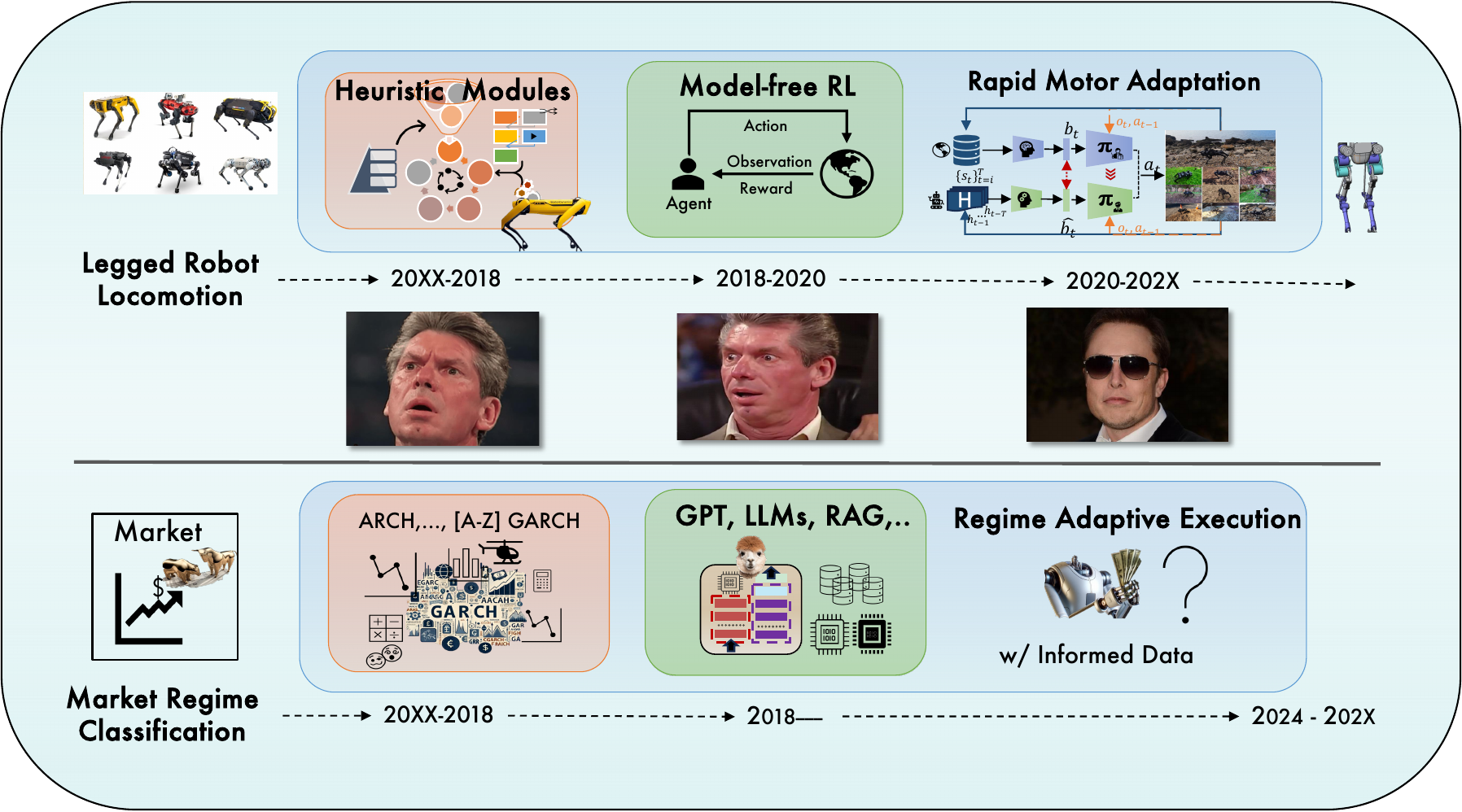}}
\vspace{0.5 em}
\caption{We propose \textbf{Regime Adaptive Execution} in the \textit{financial market} setting motivated by the success of reinforcement learning inspired robust locomotion methods supplanting intricate heuristic control architectures in quadrupedal robots, thereby eschewing decades-old conventional heuristic approaches to the  `\textit{market regime classification problem}'.}
\label{fig:timeline}
\end{center}
\vskip -0.1in
\end{figure}

At a high level, techniques from the deep reinforcement learning (RL) toolkit have enabled some of the most prominent breakthroughs in the field, from invincible game-playing agents~\cite{alphastar, alphago, alphazero}, to champion-level drone racing~\cite{kaufmann2023champion}. Examined closely, \textit{ceteris paribus}, the simulation of model interactions in environments~\cite{sim2real-tobin2017domain} plays a key enabling role in such RL-based approaches. Intuitively, simulation allows a learner agent to experience myriad combinations and feedback from an environment, thus allowing better generalization. For example, in the case of AlphaGo, the agent experiences millions of hours of game-play via `self-play' -- essentially environment interactions -- allowing it to learn superior patterns compared to traditional gaming engines. Learning in simulation then transfer to real-world works in complex domains, too. Robotics, specifically agile locomotion for quadruped robots~\cite{sim2real-peng2018sim, sim2real-tan2018sim}, best showcases its efficacy. In this domain, transferring learned skills across the \textit{sim-to-real gap} -- i.e., the environmental differences of simulated versus real world -- posed a challenge. However, more recent techniques and advancements have mostly circumvented this issue~\cite{sim2real-robotics-hwangbo2019learning, robotics-peng2020learning}. 

In this work, we juxtapose and explore the efficacy of techniques that allow robots to adapt and generalize locomotion in unseen terrains~\cite{robotics-peng2020learning, robotics-lee2020learning, robotics-kumar2022adapting} in a vastly different and more complex domain: the financial market.  

Financial market forecasting is a hard problem, arguably more complex than any of the aforementioned problems. While the allure of solving this problem transcends well-beyond the academic and scientific communities (for obvious, intrinsic rewards), the core problem can be formalized from an RL (or optimal control) perspective by setting up the problem as a (highly-complex) partially observable MDP (POMDP)~\cite{kaelbling1998planning} (see \S\ref{sec:preliminaries}). The true, plausibly large number of variables and mechanics that move the market are hidden or unobservable. Thus, reliable market simulation, thereby generating randomized market value trajectories to train agents in simulation is not yet effective, making market prediction in essence a \textit{one-shot} learning task with only one true trajectory or available environment history. Any mapping of input observations ($o_t \in \mathcal{O}$) to output price movement (i.e., market/environment reaction) learned via traditional ML techniques does not generalize well to out-of-domain (or, regime-shifted) distributions due to the hidden, 
underlying correlation and covariate shifts in a dynamic market regime~\cite{regime_ang2012regime, regime_guidolin2008size}. Basically, even if we are able to train a model that fits perfectly to past market trajectories (i.e., success in backtesting), it does not guarantee future accuracy.  

Our solution to this dynamic market regime adaptation problem is motivated and ideated by recent, remarkable successes of RL-based adaptive quadruped locomotion techniques in the robotics domain that use two-stage training of \textit{teacher-student} policies~\cite{robotics-lee2020learning, robotics-kumar2022adapting}. Bifurcating training into two stages allow (in a preliminary step) imbuing a \textit{teacher} policy with true environmental information -- termed as `\textit{privileged information}' -- that are unavailable to policies during execution. The idea is to accelerate learning in simulation with rich, denser rewards by cheating~\cite{chen2020learning}. The \textit{student} is then trained using guidance from the learned teacher, and proprioceptive signals history (without cheating). These techniques demonstrate robust zero-shot generalization to apriori unseen terrains by rapidly adapting to the environment even in absence of exteroception (i.e., blind robots), and relying only on proprioceptive sensory signals. Historically, application of techniques, motivated by the robotic pipeline, in the finance domain would be infeasible. First, we have to navigate through the bottleneck (more aptly, scale the mountain of) \textit{market simulation}. Further, we did not have anything akin to a proxy for \textit{privileged information} or, underlying environment knowledge to use -- until we did: with the meteoric advent of RL-tuned large language models (LLMs)~\cite{rlhf_instructgpt_ouyang2022training, li2023chatgpt, openai2023gpt4}, that can effectively encode `\textit{world knowledge}' to demonstrate unprecedented capabilities in downstream tasks~\cite{zhao2023survey, wei2022emergent,bubeck2023sparks} . Thus, inclusion of state-of-the-art (SOTA) LLMs in our architecture was a key enabling factor for realizing our approach. 

We adopt a similar 2-stage training, then adaptive execution (detailed in~\S\ref{sec:model}), using pre-trained LLMs as base policies that we RL-tune introducing an automatic, natural market feedback signal as auxiliary reward. We curate the largest market relevant daily news and financial technical indicators dataset for LLM supervised fine-tuning, and preferences dataset for RL-tuning. Our experiments and empirical results show that LLMs, with their imbued generic world knowledge, can support regime adaptation with continual adaptation using RL from intrinsic, natural market rewards (dubbed \textbf{RLMF}). 

\paragraph{Contributions} Our main contributions with this work are:

1. \textbf{Regime Adaptive Execution with Informed Data}: We show a novel approach of dynamic adaptation to financial market regime shifts motivated by and drawing parallels between successful RL-based robotic locomotion techniques and LLM alignment using RL and natural (intrinsic market) reward as feedback RL (RLMF).


2. \textbf{Role of LM Embeddings} in the quality of observations or news data: we perform a comprehensive analysis and discussion on the quality of source embeddings in terms of information gain from the corresponding input observations or news data. We present two interesting findings from the analysis: i. the underlying LLM architectural blocks (\textit{encoder}, \textit{decoder}, and \textit{encoder-decoder}) of models matter, and ii. model size matters, i.e., the rich gets richer since embeddings from larger  LMs (by parameter size) tend to have better \textit{information gain} compared to smaller-variants and architectures of the same family~\S\ref{app:embeddings}.

\paragraph{Organization} The rest of the paper is organized as follows:~\S\ref{sec:model} details the `RAEiD' architecture and model, \S\ref{sec:experiments} presents the experiments and results obtained, subsequent sections discuss related works~\S\ref{sec:related}, limitations and open questions~\S\ref{sec:limitations-future.works} of our work.

\section{Preliminaries and Background}\label{sec:preliminaries}
\begin{figure}[!h]
\begin{center}
\centerline{\includegraphics[width=\columnwidth]{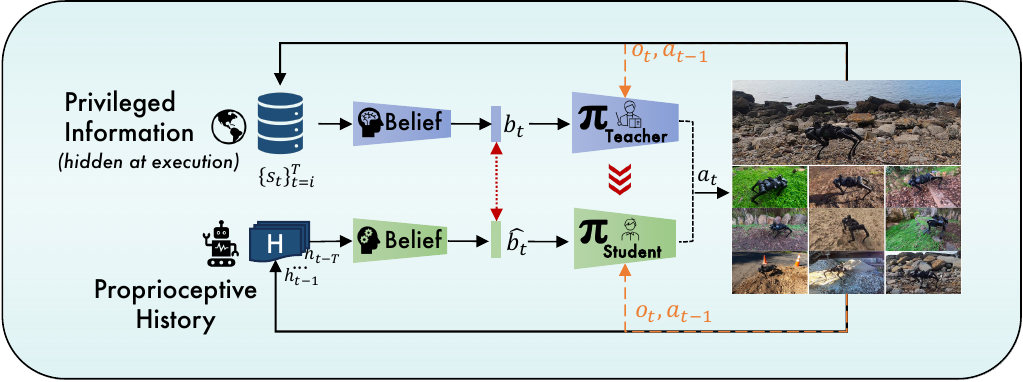}}
\vskip 0.2cm
\caption{\textbf{Robot locomotion}: high-level schematic of common dual-policy SOTA approaches.}
\label{fig:rma}
\end{center}
\vskip -0.2in
\end{figure}

\paragraph{Rapid Motor Adaptation for Quadrupedal Robots}
For a comprehensive historical background and detailed discussion on the evolution of legged robot locomotion research, we encourage the reader to any suitable survey papers, like~\cite{robotics_survey_biswal2021development}. Here, we zoom in on selected, successful methods~\cite{robotics-lee2020learning, robotics-kumar2022adapting, robotics-miki2022learning}, beginning in the 2020s, that adopted reinforcement learning (RL) in simulation for training controllers showing remarkable zero-shot generalization and demonstrated robustness in dynamic quadrupedal locomotion in challenging terrains (Fig.~\ref{fig:timeline}).  

Until circa 2018, conventional approaches to locomotion generalization on complex terrains resulted in increasingly intricate control architectures that required integration of multiple system modules. For instance, terrain mapping and map processing, motion planning, and motion control governed by separate, integrated modules with interaction dependencies during execution~\cite{robotics-old-fankhauser2018probabilistic}. Further, reliance on exteroceptive sensors, such as stereo cameras and LiDAR, were required as perceiving  and accounting for terrain properties were particularly crucial for locomotion success in complex terrains~\cite{robotics-old-bellicoso2018advances}.

The 2018-19 period saw increasingly more works and trend of applying \textbf{model-free RL} techniques to robotic legged locomotion, although mostly in experimental, lab settings~\cite{robotics-mfrl-hwangbo2019learning, robotics-mfrl-haarnoja2018learning, robotics-mfrl-xie2020learning}. Concurrently, and often in conjunction, there were surges in complementary techniques, like learning in simulation and sim-to-real transfers~\cite{sim2real-tobin2017domain, sim2real-peng2018sim, sim2real-tan2018sim}. However, starting in the 2020s, conventional controllers based on elaborate state machines and dependencies on heuristic system modules saw a radical shift towards novel approaches that used reinforcement learning in simulation to train neural networks as (locomotion) controllers~\cite{robotics-mfrl-xie2020learning} that achieved remarkable robustness and \textit{zero-shot} generalizability in unseen, challenging terrains, while acting on a stream of proprioceptive signals only~\cite{robotics-lee2020learning, robotics-kumar2022adapting}. \cite{robotics-peng2020learning} learns adaptive policies using imitation learning of real world animals. Generalization robustness was achieved by domain randomization and adaptation in simulation and sim-to-real transfer was done using advantage-weighted-regression. ~\cite{robotics-miki2022learning} extends previous works by learning from exteroceptive signals to improve performance.

Of these works, we use~\cite{robotics-lee2020learning, robotics-kumar2022adapting} for focused elucidation. While these two works each have practical nuances, at a methodology level, we found them to be mostly identical. The output action $a_t \in \mathbb{R}^D$ is the predicted joint position for $D$ robot joints (similarly, for the market prediction problem, $a_t$ is a binary or trinary (with `neutral' label) movement label prediction). Presuming a MDP formulation of the control problem, the teacher policy can be trained with any off-the-shelf RL algorithm (e.g.~\cite{trpo_schulman2015trust} under fully observable condition (access to `privileged information').  The student policy is trained via supervised learning using MLE loss (regressing) between the teacher's action prediction and their \textit{belief state} $b_t$ encoding: 

\begin{equation}
    \mathcal{L}_{\pi}(\phi) = \|\hat{x}^t_r - x_r^t\|^2 + \|\hat{b}_t(H) - b_t(x_p) \|^2
\end{equation}
where $x_r^t$ is the action ($a_t$) label at time $t$. The hat ($\hat{.}$) denote \textit{student} values, and $H$ is the proprioceptive history observed by the student, and is encoded using a temporal sequence model~\cite{tcn_bai2018empirical}, instead of prior vanilla MLP encoding of belief state -- ~\cite{robotics-lee2020learning} cites this architectural change as one of the three major ingredients of their techniques success besides `priviliged information' and adaptive curriculum learning of environment in simulation.


\begin{figure*}[!t]
\begin{center}
\centerline{\includegraphics[width=\textwidth]{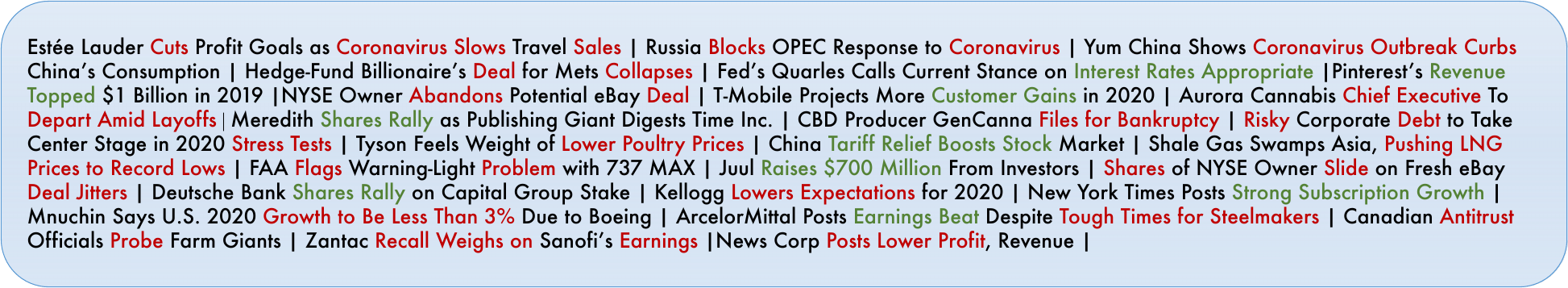}}
\caption{A snapshot of the `news` key value on date: 2020-02-06, at the upstart of the global coronavirus epidemic. Our $\pi_{LM}$ policy's prompt is composed of task instruction as query prefix, market context, and this news value concatenated: $s.t.$ 
$x_p \leftarrow (x_{instruction}; x_{context}; x_{news})$. The semantic text colors \textcolor{red}{red}, and \textcolor{green}{green} conveys negative and positive sentiments. The day's market relevent news was dominated by mostly negative sentiments. }
\label{fig:nifty-news}
\end{center}
\vskip -0.2in
\end{figure*}

\paragraph{Regime-Switching in Finance}
In empirical finance literature, regime switching processes are modeled as \textit{Markovian Switching Models}, introduced by the seminal work of Hamliton~\cite{markovswitching_hamilton1989new}, in the 1990s. The canonical regime switching problem can be presented by letting $o_t$ be an outcome variable for a market process, which recurrently depends on its own past history, $y_{t-1}$, $\varepsilon_t$ representing random shocks and (for ML/RL community, a conveniently termed) $s_t \in \{0,1,...,k\}$ a discrete random variable modeling some underlying \textit{regime process} at time, $t$. Then regimes affect the intercept(mean), $\mu_{s_t}$, auto-correlation, $\phi_{s_t}$, and volatility, $\sigma_{s_t}$, of the process~\cite{regime_hamilton2010regime}:

\begin{equation}\label{eqn:fin-regime-switching}
    o_t = \mu_{s_t} + \phi_{s_t} o_{t-1} + \sigma_{s_t} \varepsilon_t, 
                                                        \quad \varepsilon_t \sim \operatorname{iid}(0, 1).
\end{equation}

The literature and works on modern heuristic solutions to detecting, classifying, or adapting to such canonical regime switching models are too broad for our scope, and also goes against the motivation of our work (of eschewing such methods). Enthusiastic readers are encouraged to read \cite{regime_guidolin2011markov, markov_switching_hamilton1990analysis, regime_hamilton2010regime} for a detailed overview of Markovian switching models. For a comprehensive appreciation and answer to `\textit{why regime adaptation is important}?', we highly encourage reading~\cite{regime_ang2012regime, regime_guidolin2008size} -- as we have already covered the high-level intuition earlier. 

Modern deep learning based techniques essentially subsume and skip the problem of regime classification as an intermediary step to some means (like market prediction), and allow the distributional latent embeddings to encapsulate the true regime state from some input data (as a belief $b$ encoding from POMDP formulation). In essense, we too, are adhering to this paradigm, however, unlike other the other methods (relying on deep learning or RL based solutions), we dynamically adapt and update the learned policy using our proposed methodology.

\paragraph{Reward based alignment of Language Models}
Tuning pretrained LMs using reward feedback and RL enables remarkable capabilities of current chat-bots and assistants to follow instructions. The RLHF pipeline~\cite{rlhf_ziegler2019fine, rlhf_stiennon2020learning,rlhf_instructgpt_ouyang2022training} is a well-formulated approach in the NLP domain. While variants to RLHF have been proposed~\cite{dpo_rafailov2023}, we discuss only the popular RLHF pipeline for our purposes here. At a high-level, the RLHF pipeline starts with fine-tuning a pre-trained LM in supervised manner (typically with the same LM objective, but on new, high-quality domain-specific data) to obtain $\pi^{SFT}$, then training a reward model $f^{RM}_\theta$ that, once trained, is able to evaluate (usually pairs of) LM generated prompt ($x_p$) completions: $ (\hat{x}^1_r, \hat{x}^2_r) \sim \pi^{SFT}(x_p) $ and provide scalar reward $f^{RM}_\theta(\hat{x}_r) \rightarrow r \in \mathbb{R}$. A human labelled preferences dataset is typically (we deviate from in our presented approach) used to for the reward model training using MLE objective. In the final step, the domain fine-tuned LM, and the trained reward model is used to fine-tune an aligned policy using RL (e.g. PPO~\cite{ppo_schulman2017proximal}) where $\pi^{SFT}$ acts as the reference based policy: $\pi^{ref}$. PPO uses the base, reference model to impose a KL-divergence penalty during RL fine-tuning using reward feedback to ensure the fine-tuned model does not deviate or diverge too far away from the base policy and preventing unwanted scenarios like mode-collapse to high-reward answers.

\begin{figure}[ht]
\centering
    \begin{subfigure}{\columnwidth}
      \centering
      \includegraphics[width=\linewidth]{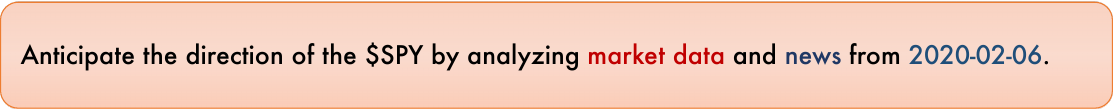}
      \caption{Instruction component of a $\pi_{LM}$ policy query $x_q$.}
      \label{fig:nifty-question}
    \end{subfigure}
\vskip 0.1in
    \begin{subfigure}{\columnwidth}
      \centering
      \includegraphics[width=\linewidth]{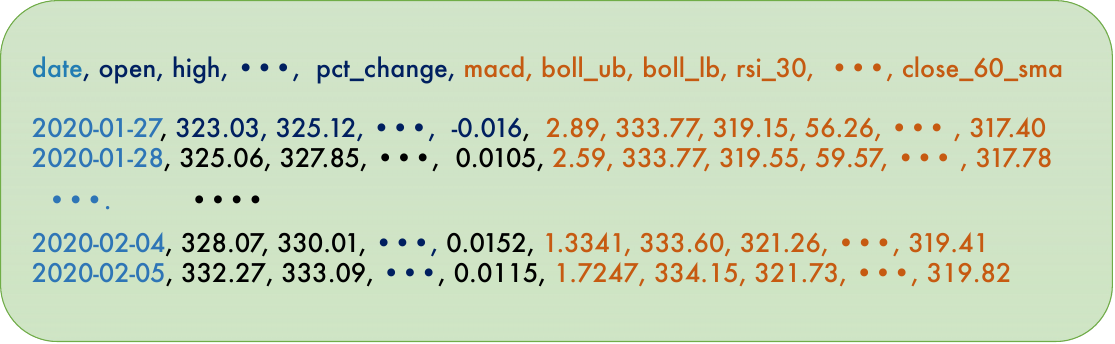}
      \caption{The market's \textbf{history} is provided as the past $t$ days of numerical statistics like the (OHLCV) price (in blue) and common technical indicators (in orange) (e.g. moving averages) data.}
      \label{fig:nifty-context}
    \end{subfigure}
\caption{Breaking down the instruction or prompt prefix, and market context components of a prompt, $x_p$.}
\label{fig:nifty-query}
\vskip -0.2in
\end{figure}

\paragraph{Problem Formulation}
Aligned with the quadrupedal locomotion and regime switching formulation, we model the task of market movement direction as a POMDP problem. We detail pertaining canonical definitions and terminology in the Appendix~\S\ref{app:ss:definitions}. 
Here, we decompose the POMDP problem as an \textbf{MDP over belief states}~\citet{kaelbling1998planning}. Thus, a policy's belief state at time $t$, $b_t$ can be seen as a sufficient statistic of the history $h_t$ towards deciding optimal actions. 

Going forward, observation at time $t$, $o_t$, will be referred to as a LM query, $x_q$ comprised of a prompt $x_{p_t}$ and action prediction label from previous time step: $\hat{x}_{r_{t-1}}$ (Fig.~\ref{fig:nifty-query}).



\section{Regime Adaptive Execution}\label{sec:model}
There are two distinct phases in our proposed approach. In the \textit{training} phase, we train a fine-tuned, and aligned language model as our \textit{teacher policy} $\pi^{teacher}_\phi$, and a \textit{reward model} $f^{RM}_\theta$, following the well-formulated RLHF pipeline~\cite{rlhf_ziegler2019fine, rlhf_stiennon2020learning,rlhf_instructgpt_ouyang2022training}, and using samples from our contributed NIFTY datasets. 

\subsection{The NIFTY Dataset}\label{ssec:nifty_dataset}
We delegate the full details of our contributed datasets NIFTY-LM ($\mathcal{D}_{LM}$) and NIFTY-RL ($\mathcal{D}_{RL}$) to Appendix~\S\ref{app:nifty-dataset}. Each JSON-object line of the $\mathcal{D}_{LM}$ contain high-quality, processed (one-turn) conversational query, where a query $x_q$ comprises of a prompt $x_p$ and a response $x_r$, i.e.,  \( x_q = (x_p; x_r) \). Thus, this dataset samples can be used for supervised fine-tuning (SFT) of a pretrained LM policy using the language modeling objective. Similarly, the \texttt{NIFTY-RL} dataset compiles a preferences dataset for rejection sampling and RL fine-tuning availing samples of chosen and rejected labels: 
$\mathcal{D}_{RL} = \left\{ \left(x^{(i)}_p, x_{r_w}^{(i)}, x_{r_l}^{(i)}\right) \right\}_{i=1}^{N}$
where ($x_{r_w} \succ x_{r_l} | x_p$).

\paragraph{Supervised Fine-tuning Teacher Policy}
The loss on a sequence \textbf{x} (comprised of tokens $x_1, ..., x_T$) from a vocabulary of size $V$ is the autoregressive cross-entropy loss (presuming a decoder-only transformer model akin to the GPT series~\cite{gpt3-brown2020language}:

\begin{equation}\label{eqn:lm-sft-loss}
    \mathcal{L}(x, \boldsymbol{\theta}) =-\sum_{t=1}^T \log P_{\hat{y} \mid x}\left(x_t \mid x_{1: t-1} ; \boldsymbol{\theta}\right)
\end{equation}
where $P_{\hat{y}\mid x}$ is the output distribution of a model parameterized by $\theta$. 

\paragraph{Training a Reward Model} 
We train a reward model $f^{RM}_\theta$, initialized with a SFT language model (using Eq.~\ref{eqn:lm-sft-loss}), sampling from $\mathcal{D}_{RM}$ in a MLE fashion formulating the preferences labels as a binary classification problem and optimizing for the negative log-likelihood loss:

{
\small
\begin{equation}\label{eqn:rm-loss}
\small
\mathcal{L}_{RM}(\theta) = -\mathbb{E}_{\substack{(x_p,x_{r_w},x_{r_l})\\ \sim \mathcal{D}_{RL}}} \left[ \log \left( \sigma \left( r_{\theta} (x, x_{r_w}) - r_{\theta} (x, x_{r_l}) \right) \right) \right]
\end{equation}
}

where $r_{\theta}(x_p,x_r)$ is a scalar reward for prompt $x_p$ and response $x_r$ with parameters $\theta$, $x_{r_w}$ is the preferred or chosen response out of the pair $(x_{r_w},x_{r_l})$ sampled from $\mathcal{D}_{RL}$ (see~\S\ref{app:nifty-dataset}).


\begin{figure}[!ht]
\vskip 0.2in
\begin{center}
\centerline{\includegraphics[width=\columnwidth]{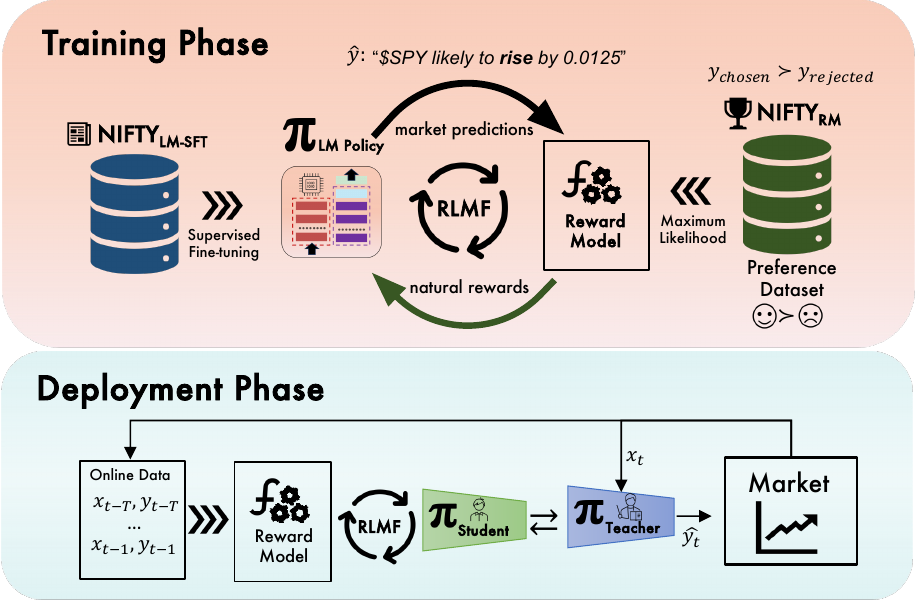}}
\vskip 0.2cm
\caption{\textbf{Regime adaptive execution} uses the NIFTY dataset to train a reward model (RM) and align a pretrained LLM during the training phase. In the deployment phase, streaming online market data is used to continually update the RM, subsequently a student policy that swaps place with an executor teacher policy after windowed intervals.}
\label{fig:arch-1}
\end{center}
\vskip -0.2in
\end{figure}

\subsection{Deriving the RLMF objective} 
\paragraph{Intuition} 
Our formulation of \textit{Reinforcement Learning from Market Feedback} or \textbf{RLMF} can be explained in a simple, intuitive manner conceptually. 
We all can formalize the market movement tomorrow based on our own beliefs (formed from our unique life-experiences or, history) about the market's state and the new information we learned today from any possible sources (news, social media, chatting with a friendly neighbour etc.). The most natural feedback or correction to our beliefs come from the true, observed movement the next day morning. However big the correction is, this feedback will (and should) not be so radical that we forget everything we have internalized from experience up until yesterday -- we are likely to attribute the mismatch to the current (likely deviated) market condition (like the inflation, war, interest rate changes etc.). 

\paragraph{Technical details}: Let $\pi^{LM}_\phi$, be a policy we want to train, that is parameterized by $\phi$. We define a policy query as: $x_q = (x_p; x_r)$. 
Let $D_{MF}$ be a dataset of size $T$ containing tuples of $(x_p, \hat{x}_r, x_r)$, where $\hat{x}_r$ is a generative completion or, response by the policy $\pi^{LM}_\phi$.
Let $f^{RM}_\theta$ be a trained reward model~(using MLE (Eq.~\ref{eqn:rm-loss})), parameterized by $\theta$. And $\pi^{ref}$ is a frozen (teacher or,) reference policy .

In our setup, the response is an action label of market movement prediction $s.t.$  $\hat{x}_r \in \{\text{rise, fall, neutral} \}$. Note that for each $\hat{x}_r$, we can collect a corresponding truth label from the market's reaction, that we denote by $x_r$. 
Having such a rollout dataset, $D_{MF}$ allows us to define a simple MLE based loss objective that we define as \textit{Reinforcement Learning from Market Feedback} (\textbf{RLMF}) loss:

\begin{align}
    \mathcal{L}_{MF}(\phi) &= \frac{1}{T}\sum_{t=1}^{T}{\|\hat{x}^t_r - x_r^t\|^2} \nonumber \\
        &= \mathbb{E}_{\substack{(x_p,x_r, \hat{x}_r) \\ \sim \mathcal{D}_{MF}}} \left[ \|\hat{x}_r - x_r\|^2 \right]
    \label{eqn:mf-loss}
\end{align}


The regular RL fine-tuning loss~\cite{rlhf_stiennon2020learning} is defined as:
\begin{align}
\mathcal{L}_{RL}(\phi) = &\mathbb{E}_{\substack{(x_p,x_r, \hat{x}_r) \\ \sim \mathcal{D}_{MF}}} \Bigg[ r_\theta(x_p, \hat{x}_r) \nonumber \\
&- \beta \log \left( \frac{\pi^{LM}_\phi (\hat{x}_r | x_p)}{\pi^{ref}(\hat{x}_r|x_p)} \right) \Bigg]
\label{eqn:rl-loss}
\end{align}
where the KL reward coefficient $\beta$ controls the strength of the KL penalty. Using the equations~\ref{eqn:mf-loss}, \ref{eqn:rl-loss}, we can maximize the following combined objective function using RL for updating policy $\pi^{LM}_\phi$:

\begin{align}
    \mathcal{L}_{RLMF}(\phi) = \mathcal{L}_{RL} &+ \gamma \mathcal{L}_{MF}(\phi)
    \label{eqn:rlmf-loss}
\end{align}
where the MF reward coefficient $\gamma$ controls the strength of market feedback reward.

The \textit{algorithms}~\ref{algo:training-phase},~\ref{algo:deployment-phase} provide high-level pseudocode for the training and deployment phases of our approach respectively as depicted in Fig.~\ref{fig:arch-1}.
 

\begin{algorithm}[!tb]
   \caption{RAEiD: Training Phase}
   \label{algo:training-phase}
    \begin{algorithmic}
        \STATE {\bfseries Step 1:} Fine-tune teacher policy $\pi^{teacher}_\phi$
        \STATE Init $\pi^{teacher}_\phi$ $\gets$ $\pi^{LM}$ assistant (e.g., Llama2-chat-7b)
        \STATE {\bfseries Input:} $D_{tr}$ (train split of NIFTY-LM), size $m$, batch $B$
        \FOR{$b = 1$ {\bfseries to} $\lfloor m/B \rfloor$}
            \STATE Init batch queries $S_B = \{\}$
            \FOR{$i = 1$ {\bfseries to} $B$}
                \STATE Sample $(x_{p_i}; x_{r_i})$ 
                \STATE Append to $S_B$
            \ENDFOR
            \STATE Update $\pi^{teacher}_\phi$ with SFT \COMMENT{ using Eq.~\ref{eqn:lm-sft-loss}}
        \ENDFOR

        \STATE {\bfseries Step 2:} Train reward model $f^{RM}_\theta$ \COMMENT{ using Eq.~\ref{eqn:rm-loss}}
        \STATE {\bfseries Step 3:} RL fine-tune $\pi^{teacher}_\phi$ using PPO~\cite{ppo_schulman2017proximal} and $\mathcal{D}_{RL}$, NIFTY-RL preferences dataset.
    \end{algorithmic}
\end{algorithm}

\begin{algorithm}[!tb]
    \caption{RAEiD: Deployment Phase}
    \label{algo:deployment-phase}
    \begin{algorithmic}
        \STATE {\bfseries Student Policy Adaptation} 
        \STATE $t \gets 0$, $T \gets \text{freq}$
        \STATE Init $\pi^{student}$ from $\pi^{teacher}$
        \STATE Repeat every $T$ steps:
        \STATE \quad Collect $\mathcal{D}_{MF} = \{(x_p,\hat{x}^r_\phi, x^{MF}_r)\}_{t=1}^{T}$
        \STATE \quad {\bfseries Step 1:} Update $f^{RM}_\theta$ \COMMENT{with Eq.~\ref{eqn:rm-loss}}
        \STATE \quad {\bfseries Step 2:} Update $\pi^{student}_\phi$ using $f^{RM}_{\theta_{\text{upd}}}$ and Eq.~\ref{eqn:rlmf-loss}
        \STATE \quad {\bfseries Step 3:} Set $\pi^{teacher} \gets \pi^{student}$, execute for $T$ 
    \end{algorithmic}
\end{algorithm}




\section{Experiments }\label{sec:experiments}
We demonstrate the efficacy of our approach using results on the NIFTY trinary (\textit{`rise', `fall', `neutral'}) stock movement prediction (SM) task, and FLARE~\cite{finma-flare-fit_xie2023pixiu} benchmark's the binary (\textit{`rise', `fall'}) SM tasks. 

\paragraph{Flare Stock Movement Datasets} This recently released financial benchmark standardizes various existing financial domain evaluation tasks (like sentiment analysis, headlines classification, NER, etc.) using consistent LM queries $x_q$ and uses the widely adopted \texttt{LM-Eval} LLM evaluation harness~\cite{eval-harness}. The three SM task datasets are: the \textbf{CIKM} datset~\cite{fin-dataset_cikm_wu2018hybrid}, \textbf{StockNet ACL}~\cite{fin-dataset_acl18_stocknet_xu2018stock}, and \textbf{BigData22}~\citep{fin-dataset_bigdata22_soun2022accurate}. Table~\ref{table:flare_sm_datasets} shows their statistics. Full benchmark details is in the appendix~\S\ref{app:flare_datasets}. 

\begin{table}[h]
\caption{Summary of Flare stock price movement datasets.}
\label{table:flare_sm_datasets}
\small
\centering
\setlength{\tabcolsep}{3pt} 
\renewcommand{\arraystretch}{1.2}
\begin{tabular}{l|c|c|c|c|c}
    \hline 
    Data                         & Stocks   & Tweets  & Days & Start Date & End Date \\
    \hline 
    \href{https://github.com/yumoxu/stocknet-dataset}{ACL18}
     & 87       & 106,271 & 696 & 2014-01-02 & 2015-12-30     \\
    \href{https://github.com/stocktweet/stock-tweet}{BigData22} & 50       & 272,762 & 362 & 2019-07-05 & 2020-06-30 \\
    \href{https://github.com/wuhuizhe/CHRNN}{CIKM18}    & 38       & 955,788 & 352 & 2017-01-03 & 2017-12-28    \\
    \hline
\end{tabular}
\end{table}

\subsection{UNReAL Results}
We name our LLM policy trained using the RAEiD methodology as \textbf{UNReAL}: \textit{Underpinning News Reward Augmented Learning in Large Language Models}. Table~\ref{table:flare_benchmark_results} shows our results on the \textbf{FLARE Benchmark}.
While the main focus was to evaluate our model against the benchmark's stock movement tasks, we ran against all tasks in the benchmark as well. 
Since \texttt{UnREAL} was trained/fine-tuned to generate short action label predictions, it was unable to solve some of the tasks that fall under non-applicable task categories: like the headline multi-label classification, NER and financial QA datasets. However, quite interestingly, its performance in sentiment classification went up dramatically (even outperforming GPT-4)! We hypothesize that this (surprising) boost may have been due to the model's exposure to signficiant swath of market news data and corresponding trinary label generation alignment, akin to the FPB task of generating a trinary prediction (`positive', `negative' or `neutral').

\begin{table*}[t]
\centering
\caption{Performance comparison of various models on the FLARE Benchmark tasks. We copy existing results from the PIXIU~\cite{finma-flare-fit_xie2023pixiu} paper for models larger than 13B parameters. We reran all models in the 7-13B parameter range on the Flare benchmark tests. Note that our model's performance on datasets corresponding to Headline, NER, FinQA, ConvFinQA are not reported (N/A), since UnREAL was aligned to output single word 3 (action) class label predictions, whereas these 4 tasks require fundamentally different tuning.}
\label{table:flare_benchmark_results}
\vskip -0.05in
\resizebox{\textwidth}{!}{
    \renewcommand{\arraystretch}{1.2} 
    \begin{tabular}{@{}l|c|cccccccccccc@{}}
    \toprule
    Dataset & Metrics          &GPT  & OPT & BLOOM         & Bloomberg & FinMA & FinMA& FinMA  & Llama-2- & Llama-2-& GPT  & UnREAL \\
            &                  &NeoX & 66B &               & GPT       & 7B    & 30B  & 7B-full& 7b        & 7b-chat  & 4    & $\sim$7b \\
    \midrule
    FPB      & Acc \(\uparrow\)& -   &   -  & -            & -         & 0.86  & 0.87 & 0.87   & 0.2918    & 0.5742   & 0.792& \textbf{0.9329} \\
             & F1              & 0.45& 0.49 & 0.50         & 0.51      & 0.86  & 0.88 & 0.87   & 0.1417    & 0.5585   & 0.795& \textbf{0.9390} \\
    FiQA-SA  & F1              & 0.51& 0.52 & 0.53         & 0.75      & 0.84  & 0.87 & 0.79   & 0.4622    & 0.7913   & -    &  0.7659 \\
    Headline & AvgF1           & 0.73& 0.79 & 0.77         & 0.82      & 0.98  & 0.97 & 0.97   & 0.5995    & 0.6161   & - & - \\
    NER      & EntityF1        & 0.61& 0.57 & 0.56         & 0.61      & 0.75  & 0.62 & 0.69   & 0.0110    & 0.1757   & 0.268& - \\
    FinQA    & EmAcc           & -   &  -   & -            & -         & 0.06  & 0.11 & 0.04   & 0.00      & 0.00     & -    & 0.00 \\
    ConvFinQA& EmAcc           & 0.28& 0.30 & 0.36         & 0.43      & 0.25  & 0.40 & 0.20   & 0.00      & 0.00     & -    & 0.00 \\
    \bottomrule
    \end{tabular}%
} 
\end{table*}

\begin{table*}[htbp]
    \centering
    \caption{Performance comparison on (zero-shot) FLARE stock price movement tasks. We only reference the results for FinMA models from~\cite{finma-flare-fit_xie2023pixiu}. We include results on latest variants of Zephyr~\cite{zephyr_tunstall2023}, and Mistral~\cite{mistral_jiang2023} in the $\sim$7B param. range.}
    \label{tab:flare-sm-results}
    \resizebox{\textwidth}{!}{%
    \renewcommand{\arraystretch}{1.2} 
    \begin{tabular}{@{}l|c|ccccccccc@{}}
    \toprule
    Dataset & Metrics           & FinMA    & FinMA     & FinMA   & Llama-2  & Llama-2  & Zephyr    & Mistral           & GPT    & UnREAL           \\
            &                   & 7B       & 30B       & 7B-full & 7b       & 7b-chat  & 7b-dpo-qlora    & 7b-instruct-v0.   & 4      & $\sim$7b         \\
    \midrule
    ACL18       & Acc \(\uparrow\)  & 0.50     & 0.49      & 0.56    & 0.5075   & 0.5145     & 0.5107       & 0.506            & 0.45   & \textbf{0.6248} \\
                & MCC               & 0.00     & 0.00      & 0.10    & 0.0206   & 0.0176     & 0.0273       & 0.0286           & 0.0379 & 0.0013          \\
                & F1                & -        & -         & -       & 0.5027   & 0.4731     & 0.5109       & 0.4979           & 0.4803 & -               \\
    BigData22   & Acc \(\uparrow\)  & 0.48     & 0.47      & 0.49    & 0.4463   & 0.5442     & 0.5108       & 0.5292           & 0.4518 & \textbf{0.6142} \\
                & MCC               & 0.04     & 0.04      & 0.01    & -0.0363  & 0.0410     & 0.0527       & 0.1212           & 0.0157 &  0.0511         \\
                & F1                & -        & -         & -       & 0.3361   & 0.5131     & 0.5109       & 0.5441           & 0.3904 &  -              \\
    CIKM18      & Acc \(\uparrow\)  & 0.56     & 0.43      & 0.53    & 0.4602   & 0.5407     & 0.5065       & 0.4995           & 0.4891 & \textbf{0.6011} \\
                & MCC               & -0.02    & -0.05     & -0.03   & -0.0041  & 0.0536     & -0.0056      & 0.0158           & 0.0090 & 0.0087          \\
                & F1                & -        & -         & -       & 0.4333   & 0.54       & 0.5082       & 0.5068           & 0.4918 & -               \\
    \bottomrule
    \end{tabular}%
    }    
\end{table*}

Table~\ref{tab:nifty-sm-results} shows the results on SM prediction task on NIFTY's test split. 

\paragraph{Discussions} The FinMA models (all variants) are unable to solve the task. This is expected, as they were fine-tuned to perform strictly binary action (prediction) label classification. Further, they base LM used (Llama-1~\cite{llama-touvron2023llama}) is comparatively old. However, UnREAL outperforms an assistant LLM in the same param. class with much newer architecture than our base LM (Llama 2~\cite{llama2-touvron2023llama} 7B), that allows much larger context size consumption compared to Llama-2. Perhaps the most surprising result was UnREAL outperforming GPT-4 by over approx. 30\% on the NIFTY and by 15-20\% points on the Flare-SM tasks! GPT-4's lacklustre performance in comparison to some other later public models also comes as a bit surprising.

\begin{table}[h]
\centering
\caption{Performance comparison of various models on the \textbf{NIFTY} Stock Price Movement Prediction Task.}
\label{tab:nifty-sm-results}
\resizebox{\columnwidth}{!}{%
    \renewcommand{\arraystretch}{1.2} 
    \begin{tabular}{@{}l|cccccc@{}}
    \toprule
    Metrics $\uparrow$   & FinMA           & Zephyr          & Llama-2-  & GPT      & UnREAL \\
                        & (all variants)  & 7b-dpo-qlora    & 7b-chat    & 4        & $\sim$7b \\
    \midrule
    Acc             & 0.2303          & 0.4132         & 0.2808     & 0.4385   & \textbf{0.7191} \\
    F1              & 0.086           & 0.4043         & 0.1841     & 0.4515   & \textbf{0.7146} \\
    \bottomrule
    \end{tabular}%
}
\end{table}

\subsection{Role of embeddings in information acquisition}

Here, we extend the discussion of model scalability (Contribution 3) and its implications on semantic clustering. Our experiments, detailed in Appendix~\S\ref{app:embeddings}, demonstrate that \textbf{larger models generate more informative embeddings}, which in turn enhance the granularity of semantic clustering. This increased granularity is particularly relevant in the financial domain, where precise interpretation of market-related news can significantly impact predictive accuracy. 
By leveraging higher-dimensional vector spaces provided by these larger models, we observe a clear increase in \textit{information gain} for market movement, location (Fig.~\ref{fig:3x3grid}), and genre tasks. These findings corroborate our hypothesis in Contribution 3~\S\ref{sec:intro} regarding the critical role of model size in semantic analysis and forecasting in finance. 

\definecolor{darkpurple}{RGB}{51, 0, 102} 
\begin{figure*}[ht]
    \centering
    \begin{subfigure}[b]{0.24\textwidth}
        \includegraphics[width=\textwidth]{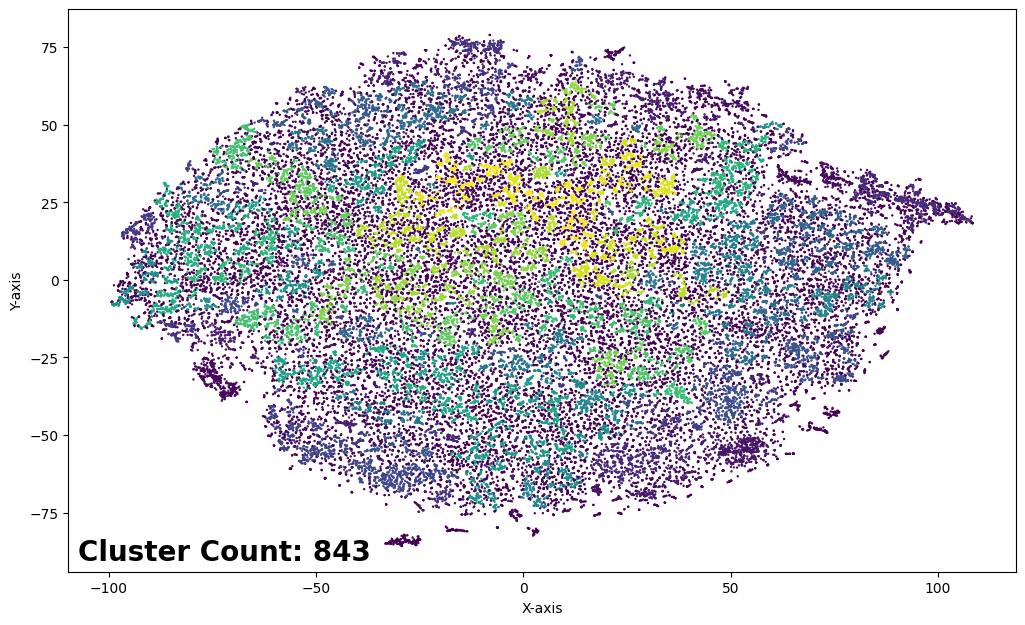}
        \caption{GPT2-SMALL}
        \label{fig:small_location}
    \end{subfigure}
    \hfill
    \begin{subfigure}[b]{0.24\textwidth}
        \includegraphics[width=\textwidth]{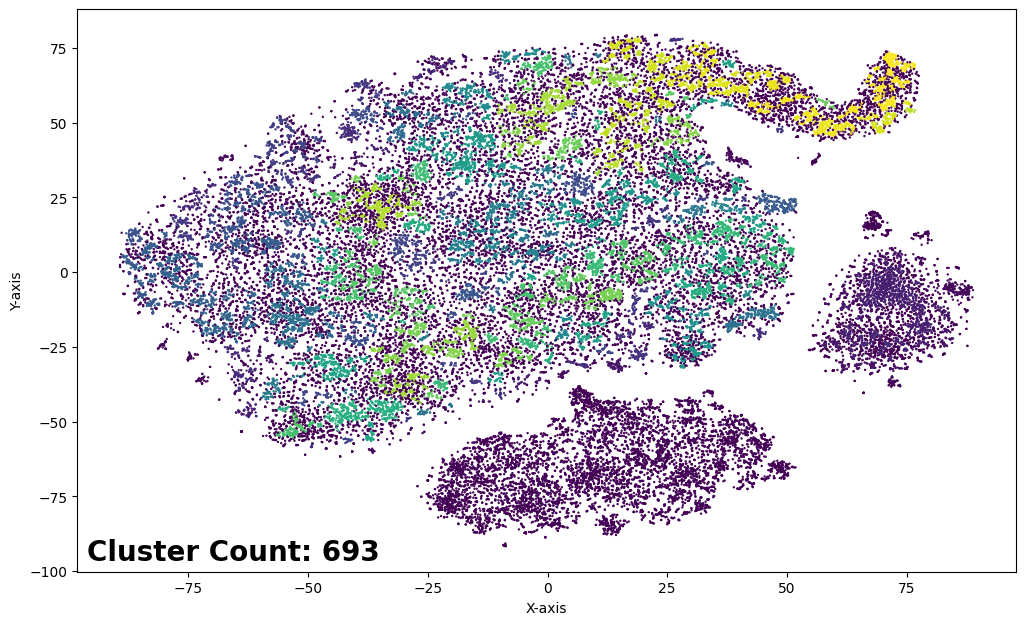}
        \caption{GPT2-MEDIUM}
        \label{fig:medium_location}
    \end{subfigure}
    \hfill
    \begin{subfigure}[b]{0.24\textwidth}
        \includegraphics[width=\textwidth]{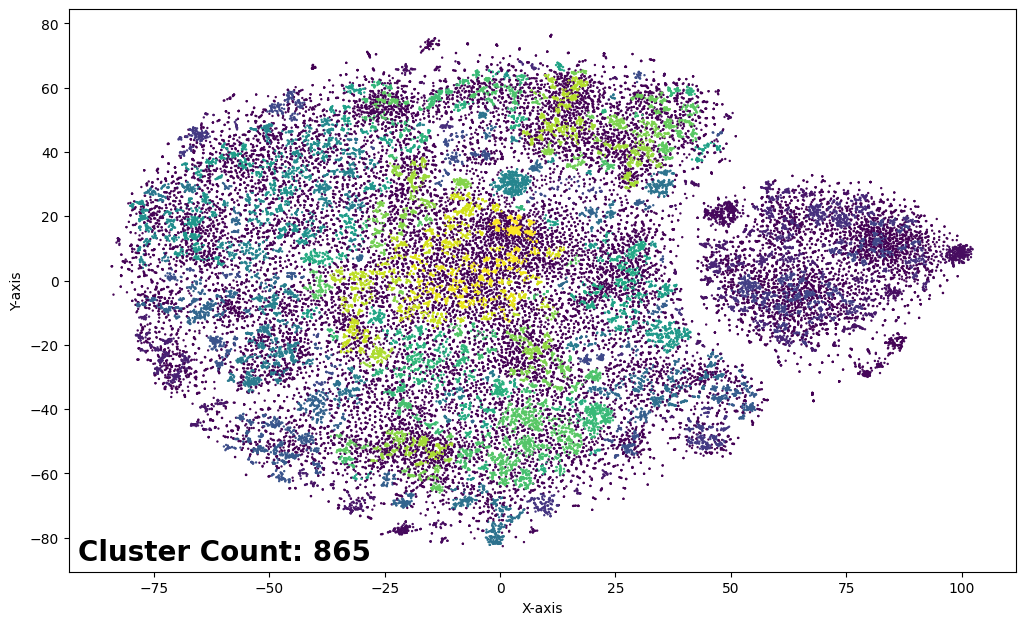}
        \caption{GPT2-Large}
        \label{fig:large_location}
    \end{subfigure}
    \hfill
    \begin{subfigure}[b]{0.24\textwidth}
        \includegraphics[width=\textwidth]{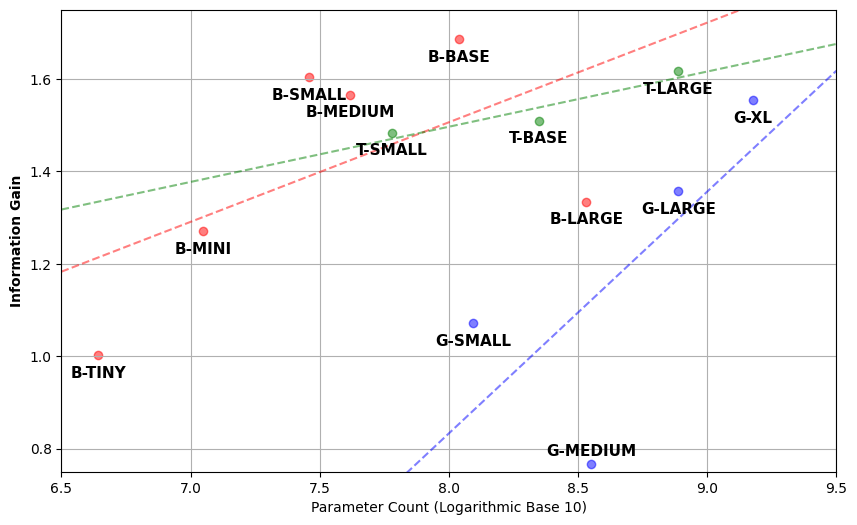}
        \caption{Location Information Gain}
        \label{fig:location_results}
    \end{subfigure}
    
    \caption{\textbf{(a-c)}: Visualizations of 2D t-SNE projections of embedded clusters (using HDBSCAN with minimum cluster size of 10) for models GPT2-[SMALL, MEDIUM, LARGE]. Each datapoint is an embedding of a news headline with a location tag in [U.S, Europe, Asia, Middle East, Latin America]. Each colour is associated with a cluster of headlines. The background purple hue are datapoints belonging to the \textcolor{darkpurple}{outlier} cluster. 
    \textbf{(d)}: \textit{Information gain} added when clustering model embeddings together on the headline location task. Information gain increases with number of model parameters. Pattern persists across model architectures: GPT2 models are shown in \textcolor{blue}{blue}, BERT models in \textcolor{red}{red}, and T5 models in \textcolor{green}{green}.}
    \label{fig:3x3grid}
\end{figure*}



\section{Related Works }\label{sec:related}
We note that our work partitions over three domains with broad, intersecting methodologies. Thus, we limit the scope of our discussion to only works closely related to ours. Fortunately, to the best of our knowledge, there aren't any prior works that directly try to juxtapose quadrupedal robot locomotion to financial market regime adaptation and movement prediction tasks. Thus, we move related works in close proximity (e.g. RL based financial trading, or LLM based market prediction tasks etc.) to the Appendix~\S\ref{app:related-work}, including history of PLMs, then LLMs in the financial domain~\S\ref{app:ss:plm-history}. 

For \textit{Natural language based financial forecasting} (NLFF), we direct interested readers to survey papers like~\cite{xing2018natural} that details recent related works. We note that while financial news has long been used for financial forecasting (e.g. most ML 101 courses focus on Kaggle's myriad market prediction tasks based on news sentiments), however, majority of such works first does (variants of) sentiment classification, i.e. attaching an (human opinionated) label of `\textit{goodness}' of the news prior to feeding that (opinionated) label for downstream forecasting, or prediction pipeline. We think such approaches are ineffective if not naive. The \textbf{sentiment} of this sentence (as we perceive it): ``\textit{The new Apple iPhones got horrendous reviews}'' is \textbf{irrelevant}; labelling (if any) should come from the market. In this case, the sentiment is positive if Apple's stock price goes up. \cite{brown2004investor}'s related work show that sentiment has little predictive power for near-term future stock returns. Further, evidence did not support the conventional wisdom that sentiment primarily affects individual investors and small stocks. \cite{antweiler2004all} explores whether Internet stock message boards can move markets.

\section{Limitations \& Future Directions}\label{sec:limitations-future.works}
\paragraph{Limitations} Firstly, we note that the goal of our work was to show the feasibility and efficacy of doing financial forecasting and regime adaptation in a fundamentally different, novel way in the current era of LLMs and AI. 
Thus, our adopted choices of LLMs -- like using Llama-2-7b~\cite{llama2-touvron2023llama} instead of larger or, newer models \cite{mistral_jiang2023, zephyr_tunstall2023}, or RL based alignment techniques instead of RL-free techniques (like DPO~\cite{dpo_rafailov2023}) etc. are perhaps best left for future works as variants sweeping for \textbf{best performance} was \textbf{not our main goal}, but showing \textbf{feasibility}/efficacy of \textbf{a new direction} was. 
Our proposed rapid adaptive execution technique is compatible/complementary to LLM architectural enhancements and variants. To elucidate, using Mistral-7B~\cite{mistral_jiang2023} as the base LLM will most likely enhance performance as it has a much larger input context length ($\simeq 8K$) compared to Llama-2 ($\simeq 4K$), thus can take in more information as query. Similarly, use of \textit{Retrieval Augmented Generation}~\cite{rag_gao2023retrieval}(RAG) techniques are complementary and compatible with our approach, and likely to enhance performance if used as a way to infuse more relevant daily financial market relevant information with the input prompt.

\paragraph{Future Directions} Our proposed approach for financial market regime adaptation, motivated by robust quadruped robot locomotion, opens up a few exciting future directions. Firstly, addressing the aforementioned limitations in the preceding paragraph to train a better performant version of RAEiDNet is a viable direction. We observe the large room for improvement on the NIFTY stock prediction task, with the current SOTA and (allegedly) trillion parameter (mixture-of-experts) model like GPT-4 achieving $\simeq46\%$ accuracy on the problem.   
Secondly, we want to point out the (deliberate) omission of any downstream financial tasks (notably, stock trading or portfolio management) in this work which we delineate as problems under the (financial) trading genre (with different evaluation metrics, like the Sharpe ratio). 
The proposed approach can be used for downstream financial tasks, including the use of RL agents performing stock trading or portfolio allocation~\cite{liu2022finrl} -- where a RAEIDNet trained adaptive policy can act as an  observation mapping module for the agent providing market regime supervision.

\section*{Broader Impact and Societal Implications} \label{app:societal_impact}
This paper presents work whose goal is to advance the field of Machine Learning. 
There can be many potential societal consequences of our work, none which we feel must be specifically highlighted here.

\paragraph{Acknowledgements and Funding}  
RS is supported by Canada NSERC CGS-D Doctoral Grant. The authors acknowledge that resources used in preparing this research were provided, in part, by the Province of Ontario, the Government of Canada through CIFAR, and companies sponsoring the Vector Institute \href{https://vectorinstitute.ai/partnerships/current-partners/}{https://vectorinstitute.ai/partnerships/current-partners/}. 

\newpage \clearpage
\bibliography{refs/main,
                refs/rl,
                refs/irl, 
                refs/qfin-datasets, 
                refs/qfin-pm,
                refs/benchmarks,
                refs/llms,
                refs/kalman-filters,
                refs/nvinden-ref-search}                

\bibliographystyle{icml2024}

\onecolumn
\appendixpage
\appendix
\section{Do Larger Models Produce Richer Embeddings?}\label{app:embeddings}

In this section, we explore whether larger models lead to richer results. When processing sentences, transformer models like GPT \cite{gpt3-brown2020language}, T5 \cite{2020t5}, and BERT \cite{kenton2019bert} produce large-dimensional vectors that capture the structure and semantic features of sentences. Consequentially, sentence embeddings localized in a group should intuitively contain more similar semantic features than those of sentences of further distance in the embedding space \cite{wieting-etal-2017-learning}. In downstream tasks, such as the market movement prediction task proposed in this paper, machine learning algorithms can then use the syntactic and semantic patterns in the embedding space to make better task-specific decisions.

We test whether larger models lead to richer embeddings by using various sizes of GPT, T5, and BERT models to create an embedding space containing news headline in the Wall Street Journal Headlines (NIFTY) dataset. We then cluster sentence embeddings that are close together, and measure whether headlines in the same cluster share similar features. The semantic similarity of these clustered features are realized by measuring the information gained after clustering embeddings. We then measure whether models with a larger number of parameters have higher clustered information gain.

\subsection{Experiments}

Testing if larger models create richer embeddings is predicated on a model's ability to group datapoints with similar features together. The features we measure are realized in \textbf{three tasks}: \textit{market movement, location}, and \textit{genre}. Each headline in the NIFTY dataset contains a single ``Tag'' that acts as a label for the category to which the headline belongs. For each task, we subsample the NIFTY, taking only task-specific tags and omitting all other rows. The tags used in the \textit{location} and \textit{genre} tasks are designed to be mutually exclusive, so a data point cannot correctly belong to two clusters. A well-performing model will create homogeneous clusters, consisting of headlines with the same tag. 

For the market movement task, we are interested in measuring how an LLM's semantic perception of a news headline can be indicative of market movement, so we only include tags relating to markets and finance. Further, in this task, we are not interested in clusters with the homogeneous tags, but instead we measure whether headlines in a cluster are indicative of homogeneous market movement. A well performing model clusters points with similar direction and magnitude of market movement from the date that each headline was published. Datasets $NIFTY_{L}$, $NIFTY_{G}$, and $NIFTY_{MM}$ are created as subsets of NIFTY with only their corresponding tags. The tags used are shown in Table~\ref{table:task_summary}.

\begin{table}[ht]
\centering
\caption{Summary of Tasks and Their Characteristics}
\label{table:task_summary}
\begin{tabular}{lcc}
\toprule
Task & Tags & Headline Count \\ 
\midrule
Market Movement & Finance, Business, Markets, Earnings & 69,068 \\
\midrule
Location & U.S, Europe, Asia, Middle East, Latin America & 49,446 \\
\midrule
Genre & Politics, World News, Business and Economy, & 382,698 \\
& Science and Environment, Health, Entertainment, \\
& Sports, Opinion and Editorial, Human Interest \\
\bottomrule
\end{tabular}
\end{table}

For each model architecture, we test multiple sizes of pretrained models, each with a different number of parameters. Each model was tested using Huggingface's transformer package \cite{huggingface_transformers}, with the exception of the OPENAI-ADA2, OPENAI-SMALL, and the OPENAI-LARGE models, whose embeddings were gathered using OpenAI's API \cite{OpenAI_API}. Parameter counts have not been disclosed for any of their embedding models, however OpenAI have noted that OPENAI-SMALL is a larger model than OPENAI-LARGE. For the T5 models, we used the small, base, and large models; and for the BERT models we used the tiny, mini, small, medium and base models. Parameter counts for each public model are available in Table~\ref{table:embedding_results}. Model's GPT2, T5, and BERT were chosen to include a decoder-only, encoder-decoder, and encoder-only model respectively.

For each model, we generated embeddings for each headline in $NIFTY_{MM}$, $NIFTY_{L}$, and $NIFTY_{G}$ datasets. Each model inputs a tokenized headline and outputs an embeddings (model embeddings are shown in ~\ref{table:embedding_results}). In order to better visualize each embedding space, we used the \textit{t-distributed Stochastic Neighbor Embedding} (t-SNE)~\cite{maaten2008visualizing} algorithm in order to reduce the dimensions of each embedding into 2 dimensions that are then plotted. t-SNE was chosen as its density-based approach outperformed principle component analysis (PCA) and uniform manifold approximation and projection (UMAP)~\cite{pearson1901liii, 2018arXivUMAP} in putting headlines into discrete clusters.

After the dimensionality of each embedding is reduced to 2 with t-SNE, we use HDBSCAN~\cite{campello2013density} to cluster our set of datapoints into discrete clusters. We require a minimum cluster size of 10 points. Datapoints that do not fit into a cluster are marked as outliers and put into their own ``outlier" cluster.

To quantify the information gain achieved through clustering, we initially computed the entropy of the unclustered multiset of tags in $NIFTY_{L}$, denoted $T_{L}$. The entropy for the base tags, $H(T_L)$, was calculated using the equation \ref{eq:Entropy}. Following clustering with HDBSCAN, we computed the total entropy of the set of clusters P, $H_C(P_T)$, using equation \ref{eq:clusteredEntropy}. Information gain associated with the clustering of location tags in described in equation \ref{eq:informationGain}, and produced $IG_{L} = H_{C}(P_T) - H(T_L)$ \cite{shannon1948mathematical}. This process is repeated for the genre tasks, using dataset $NIFTY_{G}$.

\begin{align}
p(l, T) &= \frac{|\{ i \in T : \text{label of } i = l \}|}{|T|} \label{probability} \\
H(T) &= -\sum_{l \in L} p(l, T) \log_2 p(l, T) \label{eq:Entropy} \\
H_{C}(P) &= \sum_{k=1}^{K} \frac{|P_k|}{|T|} H(P_k) \label{eq:clusteredEntropy} \\
IG &= H_{C}(P) - H(T) \label{eq:informationGain}
\end{align}

where \( L \) is a set of \( M \) tags \( (l_1, l_2, \ldots, l_M) \), \( T \) is a multiset of \( N \) tags such that each element \( t \in T \) is also in \( L \), and \( \{P_1, P_2, \ldots, P_K\} \) is a partition of \( T \) into \( K \) clusters.

For the market movement task, each headline is associated with a percent daily change in market value. Given its continuous nature, we adopted a variance-based approach as an alternative to information gain \cite{hastie2009elements}. The initial variance, $\sigma^2(T)$ (equation\ref{eq:baseVariance}), was calculated across the embeddings before clustering. Post-clustering, the variance within each cluster, $\sigma^2(P_k)$, was computed, and a weighted sum of these variances provided the overall variance after clustering, $\sigma^2_{\text{C}}(P)$ (equation \ref{eq:weightedClusterVariance}). The reduction in variance, is denoted $RV$, and is described in equation \ref{eq:infoGainVariance}. 

\begin{align}
\sigma^2(T) &= \text{Var}(T) \label{eq:baseVariance} \\
\sigma^2_{\text{C}}(P) &= \sum_{k=1}^{K} \frac{|P_k|}{|T|} \sigma^2(P_k) \label{eq:weightedClusterVariance} \\
RV &= \sigma^2(T) - \sigma^2_{\text{C}}(P) \label{eq:infoGainVariance}
\end{align}

This variance reduction approach aligns with our objective to discern the LLM's capability to semantically cluster financial news in a manner indicative of market movement. A model that is able to cluster headlines with similar percent-changes in market movement, leads to low per-cluster market movement variance, and a higher levels of information gained post-clustering.

\subsection{Results and Main Findings}

\begin{figure}[h]
    \centering
    \begin{subfigure}[b]{0.32\textwidth}
        \includegraphics[width=\textwidth]{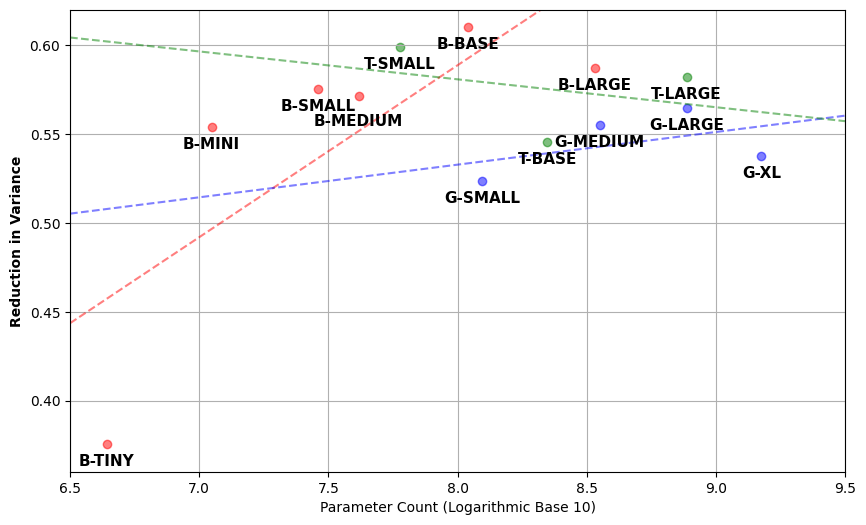}
        \caption{Market Movement Task}
        \label{fig:market_movement_results}
    \end{subfigure}
    \hfill 
    \begin{subfigure}[b]{0.32\textwidth}
        \includegraphics[width=\textwidth]{figs/nvinden_appx/location_vs_IG.png}
        \caption{Location Task}
        \label{fig:location_results_apdx}
    \end{subfigure}
    \hfill 
    \begin{subfigure}[b]{0.32\textwidth}
        \includegraphics[width=\textwidth]{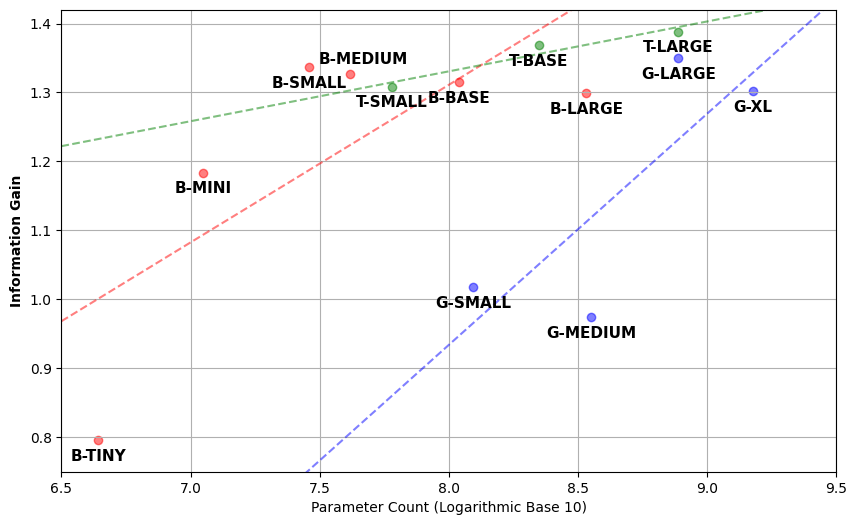}
        \caption{Genre Task}
        \label{fig:genre_results}
    \end{subfigure}
    \caption{Reduction in variance \textbf{(a)} and information gain \textbf{(b-c)} added when clustering model embeddings together on the market movement, location, and genre tasks. Multiple sizes of GPT2 (\textcolor{blue}{blue}), T5 (\textcolor{green}{green}), and BERT (\textcolor{red}{red}) models are plotted with trend line showing increase in parameter count leading to higher clustered reduction in variance and information gain. Strong correlations between parameter count and information gain are shown for all 3 model types in the location and genre tasks. In the market movement task, variance is reduced when parameter counts are increased for the GPT2 and BERT models, but not for T5 models. Although not shown in (a-c), due to having undisclosed parameter counts, OPENAI-LARGE outperformed OPENAI-SMALL in each task. All results are available in Table \ref{table:embedding_results}.}
    \label{fig:datasets}
\end{figure}

Information gain resulted from clustering our list of model's embeddings are summarized in Table~\ref{table:task_summary}, and Figure~\ref{fig:datasets}. Overall, we find that there is a \textbf{strong trend that models with a larger amount of parameters have a higher amount of information gain} in the market movement, location, and genre tasks. This leads credence to imply that larger models have the capability of creating richer embeddings on a plethora of tasks, and using larger models can lead to bigger gains in downstream tasks such as predicting market movement.

Images of subset of model clusters are available in Figure~\ref{fig:3x3grid}.

\begin{table}[ht]
\centering
\caption{Model Performance and Information Gain}
\label{table:embedding_results}
\begin{tabular}{lcccccc}
\toprule
\multirow{2}{*}{Model}  & \multirow{2}{*}{Parameter Count} & \multirow{2}{*}{Embedding Size} & \multicolumn{1}{c}{Reduction of Variance} & \multicolumn{2}{c}{Information Gain} \\ \cline{4-6} 
                        &                                  &                                 & Market Movement & Location     & Genre      \\ 
\midrule
BERT-TINY               & 4M                               & 128                             & 0.38         & 1.00         & 0.80       \\
BERT-MINI               & 11M                              & 256                             & 0.55         & 1.27         & 1.18       \\
BERT-SMALL              & 29M                              & 512                             & 0.58         & 1.60         & \textbf{1.34}       \\
BERT-MEDIUM             & 41M                              & 512                             & 0.57         & 1.57         & 1.33       \\
BERT-BASE               & 109M                             & 768                             & \underline{\textbf{0.61}}        & \textbf{1.68}        & 1.32       \\ 
BERT-LARGE              & 340M                             & 1024                            & 0.59         & 1.33         & 1.30       \\ 
\midrule
T5-SMALL                & 60M                              & 512                             & \textbf{0.60}         & 1.48         & 1.31       \\
T5-BASE                 & 222M                             & 768                             & 0.55         & 1.51         & 1.37       \\
T5-LARGE                & 770M                             & 1024                            & 0.58         & \textbf{1.61}         & \underline{\textbf{1.39}}       \\ 
\midrule
GPT2-SMALL              & 124M                             & 768                             & 0.52         & 1.07         & 1.02       \\
GPT2-MEDIUM             & 355M                             & 1024                            & 0.55         & 0.77         & 0.97       \\
GPT2-LARGE              & 774M                             & 1280                            & \textbf{0.56}         & 1.36         & \textbf{1.35}       \\
GPT2-XL                 & 1.5B                             & 1600                            & 0.53         & \textbf{1.55}         & 1.30       \\
\midrule
OPENAI-SMALL            & -                                & 1536                            & 0.45         & 1.88         & 1.33       \\
OPENAI-LARGE            & -                                & 3072                            & \textbf{0.49}         & \underline{\textbf{1.89}}         & {\textbf{1.36}}       \\
\bottomrule
\end{tabular}
\end{table}

\begin{figure}[h]
    \centering
    \begin{subfigure}[b]{0.3\textwidth}
        \includegraphics[width=\textwidth]{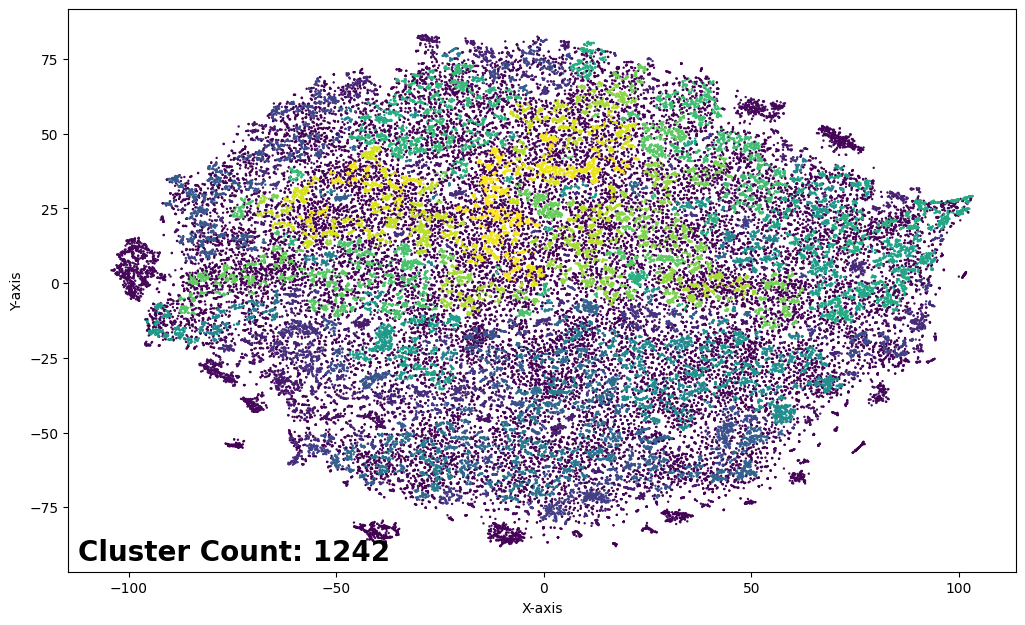}
        \caption{SMALL - Market Movement}
        \label{fig:gpt2_market}
    \end{subfigure}
    \hfill
    \begin{subfigure}[b]{0.3\textwidth}
        \includegraphics[width=\textwidth]{figs/nvinden_appx/small_location.png}
        \caption{SMALL - Location}
        \label{fig:gpt2_location}
    \end{subfigure}
    \hfill
    \begin{subfigure}[b]{0.3\textwidth}
        \includegraphics[width=\textwidth]{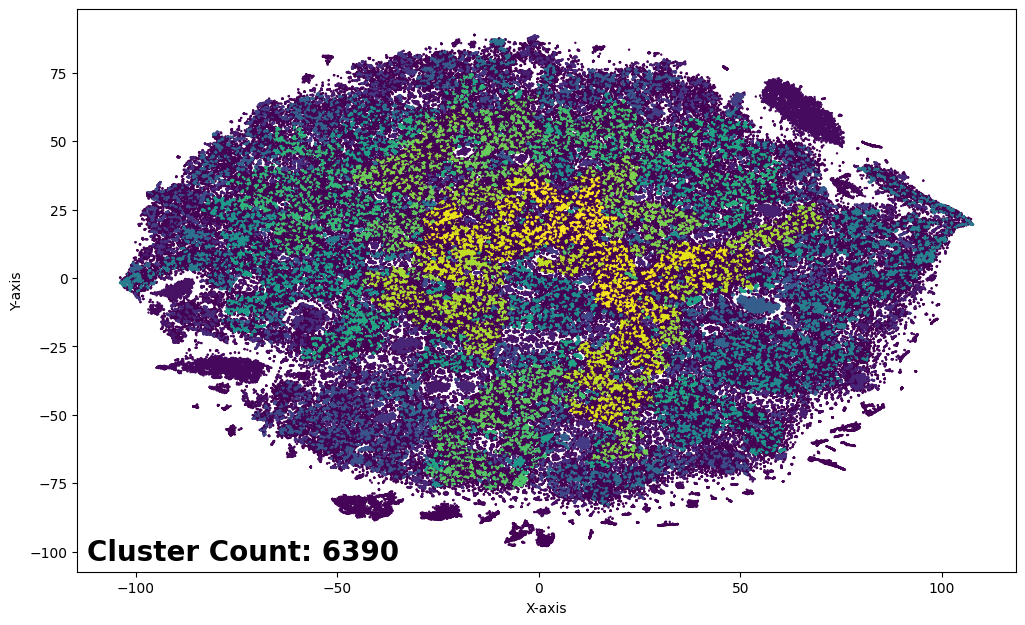}
        \caption{SMALL - Genre}
        \label{fig:gpt2_genre}
    \end{subfigure}

    \begin{subfigure}[b]{0.3\textwidth}
        \includegraphics[width=\textwidth]{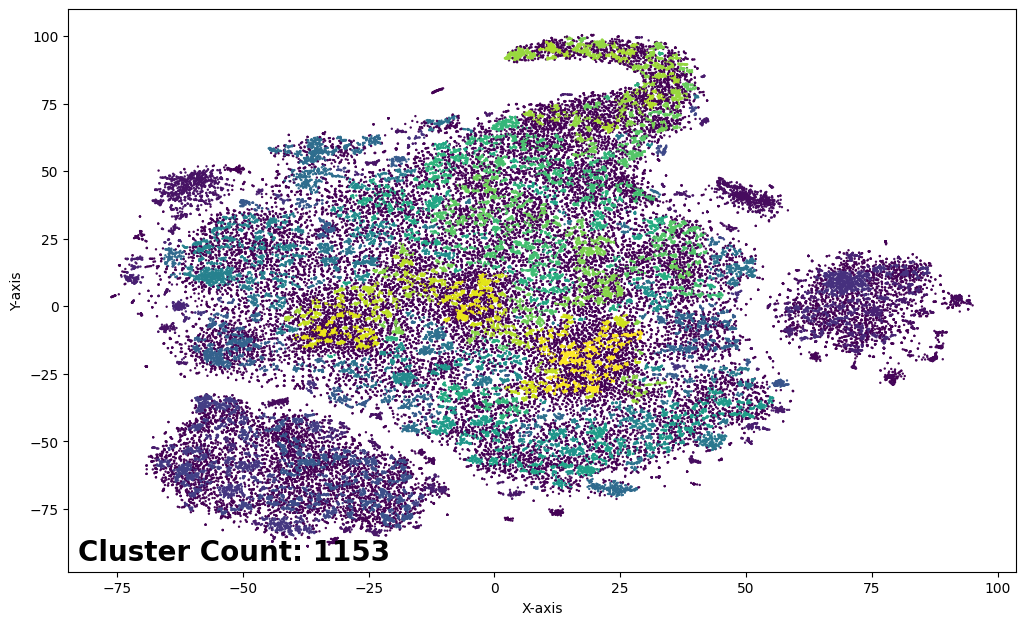}
        \caption{MEDIUM - Market Movement}
        \label{fig:small_market}
    \end{subfigure}
    \hfill
    \begin{subfigure}[b]{0.3\textwidth}
        \includegraphics[width=\textwidth]{figs/nvinden_appx/medium_location.png}
        \caption{MEDIUM - Location}
        \label{fig:small_location_apdx}
    \end{subfigure}
    \hfill
    \begin{subfigure}[b]{0.3\textwidth}
        \includegraphics[width=\textwidth]{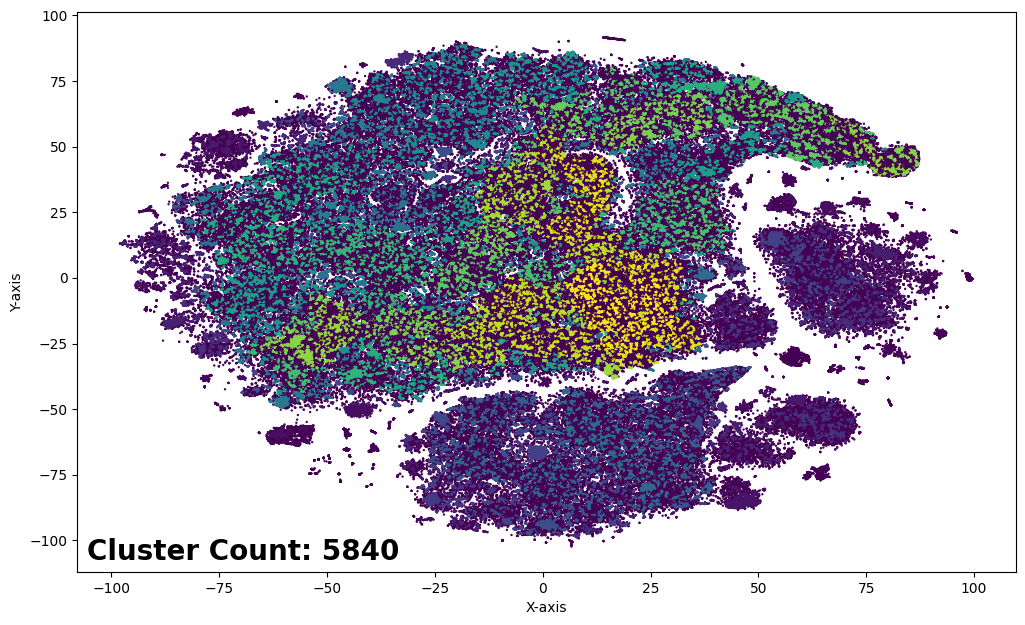}
        \caption{MEDIUM - Genre}
        \label{fig:small_genre}
    \end{subfigure}

    \begin{subfigure}[b]{0.3\textwidth}
        \includegraphics[width=\textwidth]{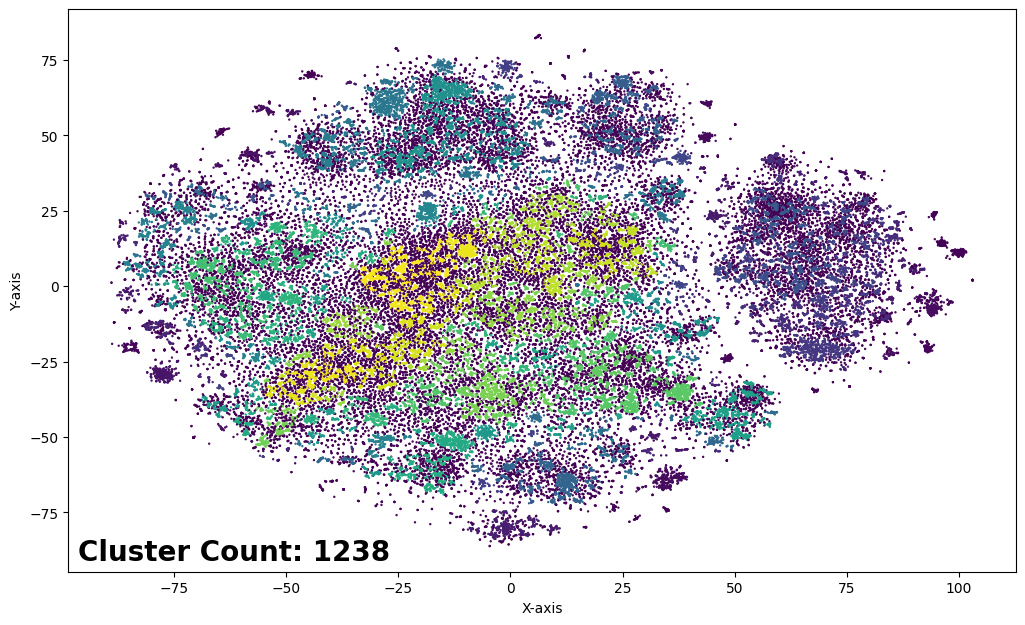}
        \caption{LARGE - Market Movement}
        \label{fig:large_market}
    \end{subfigure}
    \hfill
    \begin{subfigure}[b]{0.3\textwidth}
        \includegraphics[width=\textwidth]{figs/nvinden_appx/large_location.png}
        \caption{LARGE - Location}
        \label{fig:large_location_apdx}
    \end{subfigure}
    \hfill
    \begin{subfigure}[b]{0.3\textwidth}
        \includegraphics[width=\textwidth]{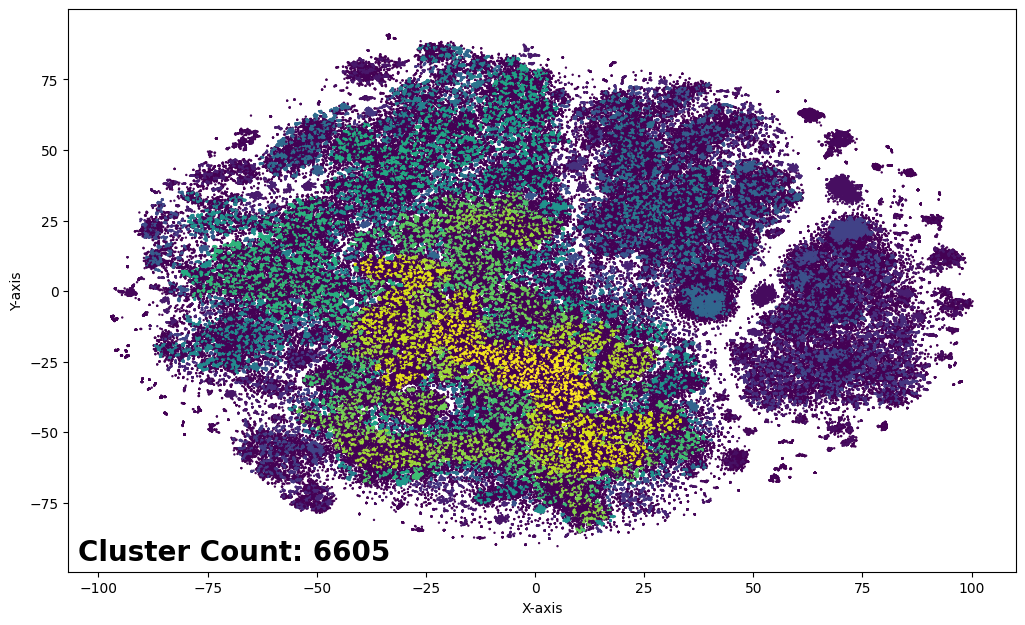}
        \caption{LARGE - Genre}
        \label{fig:large_genre}
    \end{subfigure}

    \caption{Visualizations of 2D t-SNE projections of embedded clusters on market movement, location, and genre tasks for models GPT2-SMALL, GPT2-MEDIUM, and GPT2-LARGE. Each point is a reduced and clustered headline embedding from $NIFTY$ with tags outlines in Table \ref{table:task_summary}. Each colour represent a cluster of at least 10 points. The background purple hue are points that belong to the "outlier" cluster. Results in Table~\ref{table:embedding_results} suggest larger models produce more granularity of semantic clustering.}
    \label{fig:3x3grid_apdx}
\end{figure}
\newpage
\section{NIFTY Dataset}\label{app:nifty-dataset}

The News-Informed Financial Trend Yield (NIFTY) dataset is a processed and curated daily news headlines dataset for the stock (US Equities) market price movement prediction task. NIFTY is comprised of two related datasets, NIFTY-LM and NIFTY-RL. In this section we outline the composition of the two datasets, and comment on additional details.

\subsection{ Dataset statistics }
Here, we present the statistics related to the dataset as well as other details related to the dataset. 

\begin{table}[ht]
\centering
\caption{NIFTY Dataset Statistics}
\label{table:NIFTY-stats}
\begin{tabular}{lcc}
\toprule
Category & Statistics \\
\midrule
Number of data points & 2111 \\
Number of Rise/Fall/Neutral label & 558 / 433 / 1122 \\
Train/Test/Evaluation split & 1477 / 317 / 317 \\
\bottomrule
\end{tabular}
\end{table}

\subsection{NIFTY-LM: SFT Fine-tuning Dataset}


The NIFTY-LM prompt dataset was created to finetune and evaluate LLMs on predicting future stock movement given previous market data and news headlines. 
The dataset was assembled by aggregating information from three distinct sources from January 6, 2010, to September 21, 2020. The compilation includes headlines from The \textbf{Wall Street Journal} and \textbf{Reuters News}, as well as market data of the \$SPY index from \textbf{Yahoo Finance}. The NIFTY-LM dataset consists of:

\begin{itemize}
    \item \textbf{Prompt ($x_p$)}: LLM question ($x_{question}$), market data from previous days ($x_{context}$), and news headlines ($x_{news}$).
    \item \textbf{Response}: Qualitative movement label ($x_r$) $\in{} \{Rise, Fall, Neutral\}$, and percentage change of the closing price of the \$SPY index.
    \item \textbf{Other Meta data}: Dates and data ID.
\end{itemize}


$\boldsymbol{(x_{question})}$: To generate LLM questions, we followed the self-instruct \cite{wang2023selfinstruct} framework where we used the GPT-4 model developed by OpenAI to create 20 variations of the instruction below:

\begin{quote}
Create 20 variations of the instruction below. \newline
Examine the given market information and news headlines data on DATE to forecast whether the \$SPY index will rise, fall, or remain unchanged. If you think the movement will be less than 0.5\%, then return 'Neutral'. Respond with Rise, Fall, or Neutral and your reasoning in a new paragraph.
\end{quote}

Where DATE would be substituted later, during the training phase with a corresponding date.

$\boldsymbol{(x_{context})}$: Newline delimited market metrics over the past 10 days. Each column of the dataset consists of these values. (Note: Not all market data for the past days for were available and therefore prompts might have less than 10 days of market metrics.)
\begin{itemize}
    \item \textbf{Date:} The date of the trading session.
    \item \textbf{Opening Price:} The price at which the stock opened in the market on that day.
    \item \textbf{Daily High:} The highest price at which the stock traded during the day.
    \item \textbf{Daily Low:} The lowest price at which the stock traded during the day.
    \item \textbf{Closing Price:} The final price at which the stock closed in the market on that day.
    \item \textbf{Adjusted Closing Price:} The closing price after adjustments for all applicable splits and dividend distributions.
    \item \textbf{Volume:} The total number of shares or contracts traded in a security or market during the trading day.
    \item \textbf{Percentage Change:} The percentage change in the closing price from the previous trading day.
    \item \textbf{MACD (Moving Average Convergence Divergence):} A trend-following momentum indicator that shows the relationship between two moving averages of a security’s price.
    \item \textbf{Bollinger Upper Band:} The upper boundary of the Bollinger Bands, which is typically two standard deviations above a simple moving average.
    \item \textbf{Bollinger Lower Band:} The lower boundary of the Bollinger Bands, which is typically two standard deviations below a simple moving average.
    \item \textbf{30-Day RSI (Relative Strength Index):} A momentum oscillator that measures the speed and change of price movements over the past 30 days.
    \item \textbf{30-Day CCI (Commodity Channel Index):} An indicator used in technical analysis to identify cyclical trends in a security over the past 30 days.
    \item \textbf{30-Day DX (Directional Movement Index):} An indicator that identifies the strength of a price trend over the past 30 days.
    \item \textbf{30-Day SMA (Simple Moving Average):} The average closing price of a security over the past 30 days.
    \item \textbf{60-Day SMA (Simple Moving Average):} The average closing price of a security over the past 60 days.
\end{itemize}

$\boldsymbol{(x_{news})}$: Concatenated news headlines from \textbf{Wall Street Journal} and \textbf{Reuters News}, taken from a single day.
The Reuters headlines went through a process of refinement wherein non-finance related headlines were filtered out. This filtration was achieved by performing a similarity search with SBERT model, "all-MiniLM-L6-v2" \cite{reimers2019sentencebert}.  Each Reuters headline was compared to a set of artificially generated financial headlines generated by GPT-4, with the prompt \textit{"Generate 20 financial news headlines"}. Headlines with a similarity score below 0.2, were excluded from the dataset. 

In instances where the prompt exceeded a length of 3000 words, a further refinement process was employed. This process involved the elimination of words with a tf-idf \cite{tfidf} score below 0.2 and truncating the prompt to a maximum of 3000 words.

It is also important to note that the dataset does not encompass all dates within the specified time range. This limitation is due to the absence of Reuters and Wall Street Journal headlines for certain dates.

$\boldsymbol{(x_{r})}$: Label was generated from the percentage change of the closing prices (equation \ref{eq:PCT}). They they were thresholded at values -0.5, and 0.5 to produce labels in $\{Rise, Fall, Neutral\}$ (equation \ref{eq:label}).

\begin{equation}
    \text{$PCT_{change}$} = \left( \frac{Closing Price_{t} - Closing Price_{t-1}}{Closing Price_{t-1}} \right) \times 100\%
\label{eq:PCT}
\end{equation}

\begin{equation}
    \text{$x_r$} = 
    \begin{cases}
        \text{Lower} & \text{if $PCT_{change}$} < -0.5\% \\
        \text{Neutral} & \text{if } -0.5\% \leq \text{$PCT_{change}$} \leq 0.5\% \\
        \text{Rise} & \text{$PCT_{change}$} > 0.5\%
    \end{cases}
\label{eq:label}
\end{equation}

\subsection{NIFTY-RL: Preferences Dataset}
The preference dataset is a variation of the fine-tuning dataset and it is used to train the reward model as mentioned in section \ref{sec:model}. In NIFTY-RL, labels are omitted and replaced with chosen and rejected results. The chosen result is a label corresponding to a rise, a fall or neutral movement in the stock market and is equivalent to the response in NIFTY-LM. The rejected result is a random label not equal to the chosen label.

\begin{itemize}
    \item \textbf{Prompt ($x_p$)}: LLM instruction ($x_{question}$), market data from previous days ($x_{context}$), and news headlines ($x_{news}$).
    \item \textbf{Chosen}: Qualitative movement label ($x_r$) $\in{} \{Rise, Fall, Neutral\}$. 
    \item \textbf{Rejected}: Qualitative movement label ($\Bar{x}_r$) that is randomly chosen from $\{Rise, Fall, Neutral, Surrender\} \setminus \{x_r\}$ of incorrect labels.
    \item \textbf{Other Meta data}: Dates and data ID.
\end{itemize}

\section{Additional Details}\label{app:add-details}

\subsection{Definitions and Terminology}\label{app:ss:definitions}

\paragraph{Markov Decision Process (MDP)} An MDP is defined by a tuple $(S, A, T, R, \gamma, p_0)$ where $S$ is a set of states (state space), $A$ is a set of actions, $T: S\times A \to \Pi(S)$ is the transition function, $R: S \to \mathbb{R}$ is the reward function, $\gamma \in [0, 1]$ is the discount factor, and 
$p_0: S \to [0, 1]$ 
is the distribution over initial states. A policy over an MDP is a function $\pi: S \to \Pi(A)$, and is optimal if it maximizes the expected discounted sum of rewards.

\begin{equation}
\mathcal{L} = \mathbb{E}_{\pi, T}\left(\sum_{s_i \in \tau} \gamma^i R(s_i)\right),    
\end{equation}
where $\tau = (s_0, a_0, \dots, s_T)$ is a trajectory. 

\paragraph{Partially Observable Markov Decision Process (POMDP)} 
A POMDP is a generalisation of an MDP defined by the tuple $(S, A, T, O, \omega, R, \gamma, p_0)$ where $O$ is a set of observations and $\omega: S \to \Pi(O)$ is the \textit{observation function}. An agent in a POMDP thus only receives an observation (i.e., partial information about the state) rather than the actual state of the environment. Therefore, policies on POMDPs act based on the history of observations received and actions taken at timestep $t$.

\paragraph{Belief MDPs}
Since using the complete history is impractical, many algorithms instead use  \textit{belief states} $b: O \to \Pi(S)$, which is a probability distribution over possible states updated at each timestep, given history $h_t$ comprising of previous observations. 
Intuitively, it can be thought of an agent maintaining a `belief' – a probability distribution over what it thinks the true state of the environment might be.

The belief update after taking the action $a\in A$ and receiving observation $o \in O$ is done through the following equation:
\begin{align}
    \label{eqn:belief}
    b_{o}^{a}\left(s^{\prime}\right) &= P\left(s^{\prime} \mid b, a, o\right) \nonumber \\
    &= \frac{\omega\left(s^{\prime}, o\right) \sum_{s} T\left(s, a, s^{\prime}\right) b(s)}{P(o \mid b, a)} \quad \forall s' \in S,
\end{align}
where $P(o \mid b, a)=\sum_{s^{\prime}} \omega\left(s^{\prime}, o\right) \sum_{s} T\left(s, a, s^{\prime}\right) b(s)$.

We can formulate any POMDP problem as an MDP over belief states~\citet{kaelbling1998planning}. Thus, an agent's belief state at time $t$, $b_t$ can be seen as a sufficient statistic of the history $h_t$ towards deciding optimal actions.

\subsection{FLARE Benchmark Datasets}\label{app:flare_datasets}

\begin{table}[ht]
\centering
\caption{The dataset details in the FLARE Benchmark, reproduced here from~\cite{finma-flare-fit_xie2023pixiu} as reference.}
\resizebox{\textwidth}{!}{
\begin{tabular}{@{}llrrlll@{}}
\toprule
    Data     & Task                     & Raw   & Instruction & Data Types               & Modalities    & License     \\ \midrule
    FPB      & sentiment analysis       & 4,845 & 48,450      & news                     & text          & CC BY-SA 3.0 \\
    FiQA-SA  & sentiment analysis       & 1,173 & 11,730      & news headlines, tweets   & text          & Public       \\
    Headline & news headline classification & 11,412 & 11,412      & news headlines           & text          & CC BY-SA 3.0 \\
    NER      & named entity recognition & 1,366 & 13,660      & financial agreements     & text          & CC BY-SA 3.0 \\
    FinQA    & question answering       & 8,281 & 8,281       & earnings reports         & text, table   & MIT License  \\
    ConvFinQA & question answering       & 3,892 & 3,892       & earnings reports         & text, table   & MIT License  \\
    BigData22 & stock movement prediction & 7,164 & 7,164       & tweets, historical prices & text, time series & Public       \\
    ACL18    & stock movement prediction & 27,053 & 27,053      & tweets, historical prices & text, time series & MIT License  \\
    CIKM18   & stock movement prediction & 4,967 & 4,967       & tweets, historical prices & text, time series & Public       \\ \bottomrule
\end{tabular}
}
\end{table}

\begin{table}[H]
\caption{Example prompts for the tasks in the FLARE Benchmark, reproduced here from \cite{finma-flare-fit_xie2023pixiu} as reference}
\centering
\resizebox{\textwidth}{!}{
\begin{tabular}{ll}
\hline
Data & Prompt\\
\hline
FPB & "Analyze the sentiment of this statement extracted from a financial news article. \\
    & Provide your answer as either negative, positive or neutral. \\
    & For instance, 'The company’s stocks plummeted following the scandal.' would be classified as negative." \\
\hline
FiQA-SA & "What is the sentiment of the following financial \{category\}: \\
         & Positive, Negative, or Neutral?" \\
\hline
Headline & "Consider whether the headline mentions the price of gold. \\
         & Is there a Price or Not in the gold commodity market indicated in the news headline? \\
         & Please answer Yes or No." \\
\hline
NER & "In the sentences extracted from financial agreements in U.S. SEC filings, \\
    & identify the named entities that represent a person (`PER'), an organization (`ORG'), \\
    & or a location (`LOC'). The required answer format is: `entity name, entity type'. \\
    & For instance, in 'Elon Musk, CEO of SpaceX, announced the launch from Cape Canaveral.', \\
    & the entities would be: 'Elon Musk, PER; SpaceX, ORG; Cape Canaveral, LOC'" \\
\hline
FinQA & "Given the financial data and expert analysis, please answer this question:" \\
\hline
ConvFinQA & "In the context of this series of interconnected finance-related queries and the additional information \\
           & provided by the pretext, table data, and post text from a company’s financial filings, \\
           & please provide a response to the final question. This may require extracting information \\
           & from the context and performing mathematical calculations. Please take into account the information provided \\
           & in the preceding questions and their answers when formulating your response:" \\
\hline
BigData22 & "Analyze the information and social media posts to determine if the closing price of \{tid\} \\
           & will ascend or descend at \{point\}. Please respond with either Rise or Fall." \\
\hline
\end{tabular}
}
\label{table:your_label}
\end{table}

\paragraph{FPB(Financial Phrase Bank)} Introduced by \cite{dataset-finphrasebank-malo2014good}. It contains 14,780 example sentences from finance related news which are labelled positive, negative or neutral by experts in the field.

\paragraph{FiQA-SA} Introduced by~\citep{fin-dataset_fiqa_maia201818}. It contains a total of 1,174 examples from news headlines and tweets. Each example contains the sentence and the sentence snippet associated with the target entity, aspect, and sentiment score.
An Aspect label (Level 1)  takes on one of \textit{four} possible labels (Corporate, Economy, Market or Stock), and Level 2 Aspect label takes on one of \textit{twenty-seven} possible labels (Appointment, Risks, Dividend Policy, Financial, Legal, Volatility, Coverage, Price Action, etc.). 


\paragraph{News Headline Classification} Introduced by ~\citep{sinha2021impact}. Applicable only the the commodities market, specifically gold. 

\paragraph{NER (Named Entity Recognition)} This task aims to detect and isolate crucial financial entities such as persons, organizations and locations. In the FLARE benchmark, the authors used the FIN dataset \cite{alvarado2015domain} which includes sentences from the public financial agreements through U.S. Security and Exchange Commission(SEC) filings and manually annotated entity types from LOCATION(LOC), ORGANIZATION(ORG) and PERSON(PER). (Adopted from \citealp{finma-flare-fit_xie2023pixiu})

\paragraph{FinQA} For this task, the authors use two datasets; FinQA \cite{chen2021finqa} and ConvFinQA \cite{chen2022convfinqa}. FinQA consists of Q\&A pairs annotated by experts and their corresponding earnings reports from S\&P 500 companies. ConvFinQA is a multi-turn Q\&A version of the FinQA.

\paragraph{Stock Movement Prediction Datasets and Tasks: Flare-SM tasks} 
\textbf{FLARE} proposed by \citet{finma-flare-fit_xie2023pixiu}, extends to include one financial prediction task -- the \textbf{CIKM} dataset~\cite{fin-dataset_cikm_wu2018hybrid} as an evaluation task among (four) other general financial NLP tasks. Under the hood, this benchmark is a fork of the `\textit{lm-eval}` harness~\cite{eval-harness} with addendums. Other stock price movement prediction from social dataset include \textbf{StockNet}~\cite{fin-dataset_acl18_stocknet_xu2018stock} which is mainly stock tweets of 88 stock tickers from 9 financial market industries from Twitter over two years (from 2014-2015) aligned with their corresponding historical price data. \textbf{BigData22}~\citep{fin-dataset_bigdata22_soun2022accurate} is another more recent tweets dataset comprising of tweets about 50 stock tickers during the period 2019-07-05 to 2020-06-30.

\section{Additional Related Work}\label{app:related-work}
In this section we enclose works encompassing ML/AI/RL based techniques for financial market downstream tasks, specifically tasks pertaining to market forecasting (that can be movement prediction, or, regression tasks of price forecasting).

\subsection{History of using PLMs, then LLMs in the Financial domain}\label{app:ss:plm-history}
Many PLMs for the financial domain have been proposed by continual pre-training PLMs with large-scale financial texts. \cite{araci2019finbert} proposed the first financial PLM called FinBERT that pre-trained BERT~\citep{kenton2019bert} with open released financial corpus such as TRC2financial~\citep{nist_reuters_dataset} and Financial Phrase Bank~\citep{malo2014good}. FinBERT outperforms neural network methods such as LSTM in financial sentiment classification tasks.
\cite{yang2020finbert} further proposed FinBERT by pre-training BERT with a 4.9 billion tokens financial communication corpus, which outperforms BERT on three financial sentiment classification datasets.
\cite{shah2022flue} proposed FLANG, a financial PLM with BERT and ELECTRA~\citep{clark2020electra} as the backbone. Besides English, financial PLMs in other languages, such as Chinese, were also proposed, such as Mengzi-fin~\citep{zhang2021mengzi} and BBT-FinT5~\citep{lu2023bbt}. 

\paragraph{Financial LLM Evolution}
Latest, \cite{wu2023bloomberggpt} proposed BloombergGPT, the first financial large language model with 50 billion parameters, that is pre-trained with mixed datasets from the general and financial domain. However, neither the model nor pre-trained domain datasets are released. The model is also not instruction-following like other LLMs such as ChatGPT and GPT-4.
Meta AI's LLaMA~\citep{touvron2023llama} was the first open-source LLM with parameters ranging from 7B and 13B to 65B that gained widespread traction in the research and open-source community. LLaMA-13B has comparable and even better performance than GPT-3~\citep{brown2020language} with 175B parameters on common sense reasoning tasks. Following efforts have been proposed to improve LLaMA for instruction following like ChatGPT, by instruction tuning.
Such as  the Alpaca~\cite{alpaca} model by fine-tuning LLaMA-7B with 52K instruction-following samples generated with the self-instruct method~\citep{wang2022self}. 
\cite{vicuna2023} proposed Vicuna-13B by fine-tuning LLaMA-13B with 70K conversation data from ShareGPT~\citep{sharegpt_website}. It can generate better answers to user's questions compared with Alpaca.
However, there are no open-sourced LLMs and instruction-tuning data entirely focused on the financial domain. FinMA~\cite{finma-flare-fit_xie2023pixiu} series of model along with the recently release Flare benchmark aims to fill this void, however, these models uses (Llama 1~\cite{llama-touvron2023llama}) as the base model that were not tuned to be instruction following assistants.

\subsection{More Related Works}
We enclose further works from related literature with high-level breakdown of their key contributions in each that may be of interest to target audience at the intersection of finance, RL and downstream financial tasks.

\begin{enumerate}
    \item Financial Trading as a Game: A Deep Reinforcement Learning Approach \cite{jiang2017financial}
    \begin{itemize}
        \item Stock trading AI
        \item Stock market as a dynamic environment that can be modeled as a game
        \item Deep Q Network learns to trade from market data features
    \end{itemize}
    \item A Deep Reinforcement Learning Framework for the Financial Portfolio Management Problem \cite{jiang2017deep}
    \begin{itemize}
        \item DRL to dynamically allocate funds among a set of assets
        \item MDPs to model the portfolio management task and employs a policy gradient method to optimize the investment strategy
    \end{itemize}
    \item True Knowledge Comes from Practice: Aligning LLMs with Embodied Environments via Reinforcement Learning \cite{tan2024true}
    \begin{itemize}
        \item Uses powers of LLM knowledge base, and RL's environment alignment to make better decisions
        \item novel parameter-efficient training architecture where the actor and critic share one frozen LLM equipped with low-rank adapters (LoRA) updated by PPO
    \end{itemize}
    \item Stock Market Prediction Using Deep Reinforcement Learning \cite{stockmarket-pred-awad-2023}
    \begin{itemize}
        \item Introduction of a New Framework: Proposes a combined architecture leveraging ANN, LSTM, NLP, and DRL techniques for predicting stock market trends, specifically focusing on gold stocks.
        \item Utilization of Sentiment Analysis: Employs natural language processing to process news and social media data, enhancing the prediction accuracy through sentiment analysis.
        \item Incorporation of Historical Data: Uses historical stock price data from major platforms like SandP, Yahoo, and NASDAQ to inform the predictive model.
        \item Application of LSTM and VMD: Applies LSTM networks for price prediction and Variational Mode Decomposition (VMD) for signal processing, improving prediction reliability.

        \item Innovative Use of BERT and TF-IDF: Enhances sentiment analysis phase by fine-tuning BERT models with TF-IDF for maximum accuracy in interpreting financial news sentiment.
        \item Conclusive Evidence of Efficacy: Provides conclusive results showing the effectiveness of the integrated approach in predicting stock market trends, particularly for gold stocks, with high accuracy and improved profitability potential.
    \end{itemize}
    \item LLM-Informed Multi-Armed Bandit Strategies for Non-Stationary Environments \cite{decurto2023llminformed}
    \begin{itemize}
        \item innovative strategy for the multi-armed bandit problem in dynamic environments by integrating large language models (LLMs) to guide decision-making.
        \item nparameter-efficient architecture combining LLMs with reinforcement learning to optimize the balance between exploration and exploitation.
    \end{itemize}
    \item Temporal Data Meets LLM -- Explainable Financial Time Series Forecasting(\cite{xinli-yu-temporal-data-meets-llm})
        \begin{itemize}
        \item Introduction to LLM in Finance: Investigates LLMs' capability for explainable financial forecasting, addressing challenges like cross-sequence reasoning and multi-modal signal integration.
        \item Methodology: Utilizes NASDAQ-100 stock data, company metadata, and economic/financial news for LLM-based forecasting, employing GPT-4 and Open LLaMA models.
        \item Experiments with GPT-4 and Open LLaMA: Demonstrates zero-shot/few-shot inference and fine-tuning techniques to enhance forecasting performance.
        \item Superior Performance Over Traditional Models: Shows that LLM approaches, particularly GPT-4 with Chain of Thought (COT), outperform traditional ARMA-GARCH and gradient-boosting tree models in accuracy and explanation quality.
        \item Future Directions: Suggests further research into extending studies to other stock indexes, integrating more data types, and exploring fine-tuning of larger LLMs for enhanced reasoning capabilities.
    \end{itemize}
    \item Unveiling the Potential of Sentiment: Can Large Language Models Predict Chinese Stock Price Movements?\cite{zhang2023unveiling}
    \begin{itemize}
        \item Benchmark and Framework Development: The authors introduce a comprehensive benchmark and a standardized back-testing framework to objectively assess the performance of various LLMs in extracting sentiment factors from Chinese financial news texts.
        \item Model Comparison: Three types of models are compared: generative LLM (ChatGPT), Chinese language-specific pre-trained LLM (Erlangshen-RoBERTa), and financial domain-specific fine-tuned LLM classifier (Chinese FinBERT).
        \item Sentiment Extraction and Trading Strategy: The study involves extracting sentiment factors from a large volume of Chinese news summaries and constructing quantitative trading strategies to evaluate the models' performance through back-tests.
        \item Results: The Erlangshen-RoBERTa model outperforms the others in terms of annual return, risk-adjusted return, and excess return, demonstrating the importance of language-specific pre-training and fine-tuning in sentiment analysis for the Chinese stock market.
        \item Conclusions: The research highlights the potential of LLMs in enhancing quantitative stock trading strategies by leveraging sentiment analysis, emphasizing the effectiveness of language-specific models and methodologies over general model size for Chinese financial texts.
    \end{itemize}
    \item Reinforcement Learning for Optimizing RAG for Domain Chatbots \cite{kulkarni2024reinforcement}
    \begin{itemize}
        \item The paper presents a method to optimize Retrieval Augmented Generation (RAG) for domain chatbots by using Reinforcement Learning (RL) to reduce the number of tokens required from a Large Language Model (LLM), thus saving costs while maintaining or slightly improving accuracy.
        \item It introduces a policy-based model that decides whether to fetch FAQ context for a query or not, demonstrating significant cost savings (~31percent) and improved retrieval accuracy through experimental results.
    \end{itemize}
\item TradingGPT: Multi-Agent System with Layered Memory and Distinct Characters for Enhanced Financial Trading Performance
\cite{li2023tradinggpt}
\begin{itemize}
    \item Introduces a multi-agent framework utilizing Large Language Models (LLMs) with layered memories to improve financial trading decisions, aligning closer to human memory processes.
    \item Proposes a novel method where trading agents are equipped with individualized characters and risk preferences to diversify trading strategies and enhance market opportunity identification.
    \item Incorporates real-time multi-modal data processing for comprehensive financial analysis, enabling agents to adapt quickly to market changes for both daily and high-frequency trading.
    \item Details the system architecture, including memory formulation based on individual and inter-agent experiences, and the design of training and testing workflows to optimize trading strategies.
    \item Demonstrates potential for superior trading performance through the simulation of realistic trading scenarios, aiming for future application in various domains beyond finance, like gaming and healthcare.
\end{itemize}
    
\end{enumerate}


\end{document}